\documentclass[10pt,conference]{IEEEtran}
\IEEEoverridecommandlockouts

\usepackage[T1]{fontenc}
\usepackage[utf8]{inputenc}
\usepackage{cite}
\usepackage{amsmath,amssymb,amsfonts}
\usepackage{graphicx}
\usepackage{array}
\usepackage{academicons}
\usepackage{xcolor}
\usepackage{stfloats}
\usepackage{url}
\usepackage{verbatim}
\usepackage{xspace}
\usepackage{comment}
\usepackage{csquotes}

\usepackage{amsmath,amsfonts}
\usepackage{amsthm}
\usepackage{bm}

\usepackage{amsthm} 
\usepackage{algorithm}

\usepackage{placeins}

\usepackage{algpseudocode}
\usepackage{textcomp}
\usepackage{listings}
\usepackage{color}
\usepackage{wrapfig}

\usepackage{soul}

\usepackage[most]{tcolorbox} 

\usepackage{adjustbox}
\usepackage{balance}

\usepackage{array}
\usepackage{ragged2e}
\newcolumntype{P}[1]{>{\RaggedRight\hspace{0pt}}p{#1}}
\newcolumntype{X}[1]{>{\RaggedRight\hspace*{0pt}}p{#1}}

\usepackage{orcidlink}

\usepackage{xspace}

\usepackage{amsmath,amsfonts}

\usepackage{wrapfig}

\usepackage{soul}
\usepackage{balance}
\usepackage{hyperref}
\usepackage{textcomp}
\usepackage{listings}

\usepackage{multirow}

\definecolor{codegreen}{rgb}{0,0.6,0}
\definecolor{codegray}{rgb}{0.5,0.5,0.5}
\definecolor{codepurple}{rgb}{0.58,0,0.82}
\definecolor{backcolour}{rgb}{0.95,0.95,0.92}

\usepackage{graphicx} 
\usepackage{caption,subcaption}
\usepackage{csquotes}

\usepackage{lipsum} 

\usepackage[most]{tcolorbox} 

\theoremstyle{definition}

\usepackage{pgfplots}
\usepackage{subcaption}

\newcommand{\eg}{{e.g.,}\xspace}

\newcommand{\etal}{{et~al.}\xspace}
\newcommand{\ie}{{i.e.,}\xspace}

\newcommand{\ci}{{\it (i) }}
\newcommand{\cii}{{\it (ii) }}
\newcommand{\ciii}{{\it (iii) }}
\newcommand{\civ}{{\it (iv) }}

\newcommand{\ca}{{\it (a) }}
\newcommand{\cb}{{\it (b) }}
\newcommand{\cc}{{\it (c) }}

\makeatletter
\def\@IEEEsectpunct{.\ \,}
\def\paragraph{\@startsection{paragraph}{4}{\z@}{1.5ex plus 1.5ex minus 0.5ex}%
{0ex}{\normalfont\normalsize\bfseries}}
\makeatother

\definecolor{notecolor}{rgb}{0.8,0,0} 

\theoremstyle{definition}
\newtheorem{example}{Example}

\IEEEpubid{\makebox[\columnwidth]{ 979-8-3503-7128-4/24/\$31.00~\copyright2024 IEEE\hfill} \hspace{\columnsep}\makebox[\columnwidth]{ }}

\begin{document}

\title{Investigating Timing-Based Information Leakage in Data Flow-Driven Real-Time Systems}


\author{
\IEEEauthorblockN{
Mohammad Fakhruddin Babar\IEEEauthorrefmark{1},
Zain A. H. Hammadeh\IEEEauthorrefmark{2},
Mohammad Hamad\IEEEauthorrefmark{3}, and
Monowar Hasan\IEEEauthorrefmark{1}
}
\IEEEauthorblockA{\IEEEauthorrefmark{1}School of Electrical Engineering \& Computer Science, Washington State University, Pullman, WA, USA\\
Emails: m.babar@wsu.edu, monowar.hasan@wsu.edu}
\IEEEauthorblockA{\IEEEauthorrefmark{2}Institute of Software Technology, German Aerospace Center (DLR), Braunschweig, Germany\\
Email: zain.hajhammadeh@dlr.de}
\IEEEauthorblockA{\IEEEauthorrefmark{3}Department of Computer Engineering, Technical University of Munich, Munich, Germany\\
Email: mohammad.hamad@tum.de}
}


\maketitle
\IEEEpeerreviewmaketitle

\thispagestyle{plain}
\pagestyle{plain}


\begin{abstract} 
Leaking information about the execution behavior of critical real-time tasks may lead to serious consequences, including violations of temporal constraints and even severe failures. We study information leakage for a special class of real-time tasks that have two execution modes, namely, \textit{typical} execution (which invokes the majority of times) and \textit{critical} execution (to tackle exceptional conditions). The data flow-driven applications inherit such a multimode execution model. 
In this paper, we investigate whether a low-priority ``observer'' task can infer the execution patterns of a high-priority ``victim'' task (especially the critical executions). We develop a new statistical analysis technique and show that by analyzing the response times of the low-priority task, it becomes possible to extract the execution behavior of the high-priority task. We test our approach against a random selection technique that arbitrarily classifies a job as critical. We find that correlating the observer's response times with the victim's jobs can result in higher precision in identifying critical invocations compared to a random guess. We conduct extensive evaluations with systemically generated workloads, including a case study using a UAV autopilot (ArduPilot) taskset parameters. We found that our inference algorithm can achieve relatively low false positive rates (less than $25\%$) with relatively low footprint (1 MB memory and 50 ms timing overhead on a Raspberry Pi 4 platform). We further demonstrate the feasibility of inference on two cyber-physical platforms: an off-the-shelf manufacturing robot and a custom-built surveillance system.
\end{abstract}


\section{Introduction}

Real-time systems are essential in safety-critical applications, including avionics, power grids, autonomous vehicles, manufacturing, unmanned aerial vehicles (UAVs), and healthcare~\cite{rt_autonomous,rt_manufacturing,rt_power_grid,rt_healthcare,rt_uav}.  Many data-driven real-time tasks operate in two execution modes: \ca \textit{typical} and \cb \textit{critical}. For example, in a healthcare monitoring system, a real-time task functions in typical mode to monitor stable vital signs with minimal processing but switches to critical mode upon detecting abnormal readings, such as a sudden drop in heart rate. In critical mode, the task runs complex algorithms and triggers alerts with longer execution times to ensure patient safety. Likewise, in an automated assembly line, the control system operates in typical mode for routine tasks such as monitoring conveyor speeds and sensors with standard processing. During anomalies, such as machinery malfunctions, it may switch to critical mode to perform complex diagnostics and emergency responses with longer execution times. 

\begin{figure*}[!t]
\centering
\includegraphics[scale=0.55]{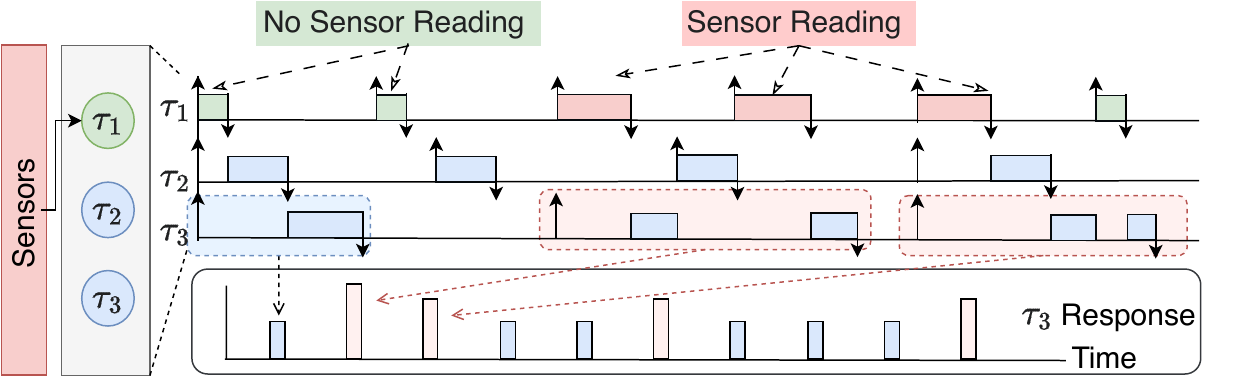}
\caption{An example of a real-time system where a task has two execution modes. The victim task $\tau_{1}$ operates in typical and critical modes. In typical mode, it runs without sensor data, while in critical mode, it processes sensor data. The observer task $\tau_{3}$ analyzes its own response time to predict $\tau_{1}$’s execution pattern.}
\label{fig:data-flow example} 
\end{figure*}

Consider a well-known application scenario where a task implements {\it Kalman filter}, as depicted in Fig.~\ref{fig:data-flow example}. The periodic task runs for a typical (short) execution time when the sensor data are unavailable, known as {\it priori}. Critical (long) execution occurs when the sensor data is available, and this execution is known as {\it potseriori}. A similar task model is also widely used in space systems\cite{Bird2001,Hammadeh2019}, recovery mechanisms in safety-critical systems\cite{hammadeh_et_al:LIPIcs.ECRTS.2017.17}, and various cyber-physical systems where operational demands change dynamically. In such environments, tasks typically run in typical mode during regular operation but switch to critical mode when responding to urgent events or specific sensor readings.

Many real-time systems are safety-critical, meaning any failure could cause serious harm to individuals, the system itself, or the environment. Over the past decade, cyber-physical systems with real-time properties have faced a growing number of attacks~\cite{attack_on_power_grid,attack_on_cps,attack_on_drone,attack_on_uav,attack_on_cps_2}. Successfully attacking such systems requires detailed system-specific knowledge, which varies based on the attack type and target component. Many real-time systems rely on periodic tasks and use deterministic schedules, making them prime targets for attackers~\cite{scheduleak,yoon2016taskshuffler}. Stealthy information leaks can lead to side-channel attacks, where adversaries exploit non-functional system attributes—such as timing variations or power consumption—to extract sensitive information covertly~\cite{covertly_1,covertly_3,covertly_2}. Preventing unauthorized information flow is crucial in safety-critical environments like avionics and industrial control. However, the predictable nature of real-time schedulers can unintentionally create opportunities for information leakage in priority-driven systems.

This research explores the potential for information leakage in data-driven real-time systems where tasks have both typical and critical execution modes. A task may perform routine operations in typical mode and switch to critical mode in specific scenarios or when handling sensitive data. An adversary could exploit such leaks (e.g., by predicting when a critical task is about to execute) to manipulate the system and cause harmful effects. Our study aims to identify information leakage in real-time systems, especially \textit{\textbf{whether an attacker can predict the future arrivals of targeted (victim) task's critical mode}}. This information is crucial as it can lead to severe consequences. For example, if an attacker can determine when a critical task is running, they could gather side-channel data (e.g., cache usage) or launch denial-of-service (DoS) attacks to block the task from executing~\cite{scheduleak}. \figurename~\ref{fig:data-flow example} illustrates an example where low priority observer task ($\tau_3$) attempts to extract information about the execution of the high-priority victim ($\tau_1$).

In this paper, we present a statistical inference model to analyze information leakage in nondeterministic dual-mode real-time systems. We show that the response time of the low-priority rogue task can be used as a leakage signal to predict the future critical arrival of the high-priority victim task. Such information leakage allows an unprivileged, low-priority task~(\ie observer task) to infer the timing behavior of a critical high-priority task (\ie victim task) by calculating its own response time. 

We made the following contributions in this research.

\begin{itemize}
     \item We analyze the problem of information leakage in dual-mode fixed-priority real-time systems (Section~\ref{sec:infleak}).
     \item We propose a statistical analysis that allows low-priority tasks to infer execution patterns of a high-priority task (Section~\ref{infer_pst}).
\end{itemize}

We perform extensive design-space exploration to assess the information leakage parameters (Section~\ref{sec:eval}). We study the overheads of the proposed inference process on a Raspberry Pi~\cite{Raspberrypi} platform (Section~\ref{sec:overhead_pi}). We further demonstrate the efficacy and consequences of the explored leakage channel on two cyber-physical platforms: an off-the-shelf robot arm and a custom-built surveillance system (Section~\ref{sec:cps_demo}).

\section{Model and Assumptions}


\subsection{System Model}

We consider a uniprocessor real-time system with $n$ periodic tasks $\Gamma=\{\tau_1, \cdots, \tau_n\}$ scheduled by a fixed-priority policy. In this work, we want to investigate whether a low-priority task (called \textit{observer}, denoted by $\tau_o$) can infer the behavior of a targeted, higher-priority one (called \textit{victim}, denoted by $\tau_v$).

Let us donate $hp(\tau_i)$ as the set of tasks with higher priority than $\tau_i$. By assumption, $hp(\tau_o) = \{ \tau_1, \tau_2,\cdots,\tau_v,\cdots \}$ and $\tau_1$ is the highest-priority task. Each task $\tau_i$ generates an infinite sequence of \textit{jobs} and is characterized by a tuple $(C_i, T_i, D_i)$. The first parameter, $C_i$, denotes the worst-case execution time (WCET).
The jobs of $\tau_i$ are periodically invoked with an inter-arrival time (period) $T_i$ and completed before the deadline $D_i$. We assume the tasks follow the implicit deadline model ($D_i=T_i$), \ie the tasks must be completed before their next periodic arrival. 
The execution times are data-driven. Hence, depending on input or processing data, each task $\tau_i$ has two execution modes: \ca typical execution time, $C_i^{typ}$ (which is the execution times during normal conditions), and \cb critical execution time, $C_i^{cri}$, that the runtime of jobs that occur depending on data flow or certain event processing~\cite{quinton2012formal}. Hence, $C_i \in \{C_i^{typ}, C_i^{cri}\}$. 
The likelihood of $C_i^{cri}$ occurring is usually lower than that of $C_i^{typ}$ and the critical execution times are generally greater than the typical runtime (\ie $C_i^{cri} \geq C_i^{typ}$)~\cite{quinton2012formal}. 

A low-priority task may delay a high-priority task if they are using the same shared resource. We represent this blocking delay as $B_i = \max_{\tau_l \in lp(\tau_i)} \{ C_l \}$, where $lp(\tau_i)$ is the set of tasks with a priority lower than $\tau_i$. Which jobs of a task $\tau_i$ will result in typical or critical execution is not known apriori. The observer task $\tau_o$ aims to predict the execution behaviors (\ie which jobs will contribute to critical invocations in the future). The leaked information may aid in launching other serious attacks (\eg overriding or blocking an actuation signal~\cite{scheduleak}). 

We note that the task model considered here is different than mixed-criticality systems~\cite{mixed_criticality_survey_alan_burns}. Unlike mixed-criticality (or dual-criticality) models where tasks have criticality levels and lower-criticality ones are dropped in overloaded situations, we consider an application-driven scenario where different instances of tasks contribute varying execution times, \eg $C_i^{typ}$ or $C_i^{cri}$, where $C_i^{typ}$ being the dominant one. Hence, from an attacker's point of view, such a system contains more ``noisy observations'' than classical real-time workload models.


This paper focuses on analyzing information leakage in non-deterministic time-critical systems, \ie whether an attacker (\eg observer task) can infer the runtime behavior of the victim. How this information can be used to carry out further attacks is \textit{not} within the scope of the work. However, we do present instances of exploring the leakage information in our demonstrations (Section~\ref{sec:cps_demo}).

\subsection{Threat Model} \label{sec:threat_model}

We consider a noninterference model~\cite{fixed_priority_sched}, \ie the scheduling parameters, the scheduling algorithm, and the resulting schedule are publicly available to the adversary. Besides, all tasks have access to precise clocks (\ie system timer). We assume that the attacker can compromise a low-priority task (\ie the observer task, $\tau_o$) to infer the arrival modes of a higher priority target (\ie the victim task, $\tau_v$).  We do not have any specific assumptions on how the attacker task ($\tau_o$) can be modified to inject inference logic. 
 For instance, the adversary can use known system vulnerabilities or zero-day bugs, exploit supply-chain integration, or apply social engineering tactics~\cite{yoon2017learning,isorc_23_mh}. This is a typical assumption in existing real-time security literature~\cite{yoon2016taskshuffler,scheduleak,hasan_contego,tcps_22_contego}. 
 We focus on analyzing information leakage in non-deterministic time-critical systems and study whether a malicious task (\ie $\tau_o$) can infer the runtime behavior of a high-priority task (\ie $\tau_v$). 
 While how the leaked information is exploited by the adversary is not the primary focus of this investigation, we do present instances of exploiting the leakage information through our cyber-physical testbed demonstrations (see Section~\ref{sec:cps_demo}).

\section{Information Leakage in Periodic Priority-Driven Systems} \label{sec:infleak}


Consider a time interval of $D_i = T_i$. In the worst case, $\tau_i$ will be preempted (\ie interfered) $\lceil\frac{T_i}{T_h}\rceil$ times by the jobs of a high-priority task $\tau_h$ with inter-arrival time $T_h$. Hence, total  interference delay for $\tau_i$: $I_i = \sum_{\tau_h \in hp(\tau_i)}\lceil\frac{T_i}{T_h}\rceil C_h$. Adding interference and blocking delay with the task's execution time provides the response time (\ie $R_i = C_i + B_i + I_i$). By assumption, a set of tasks is \textit{schedulable} if $R_i \leq D_i, \forall \tau_i \in \Gamma$.

We show that it is feasible to infer non-deterministic task arrival behavior (say typical and critical jobs) by observing the impact that the low-priority task has on its own response time (\ie the time between arrival and completion of a task).
Based on this simple analysis, we show that there exists a correlation between the victim's execution of typical/critical jobs and the observer's response times. Let us consider a taskset presented in Table~\ref{tab:example_data_1}, where there is no blocking delay (\ie $B_i = 0$ for all three tasks). The victim's typical and critical execution time is 1 unit and 3 units, respectively, and the observer's both typical and critical execution time is 2 units. Depending on the $\tau_v$'s job execution modes (typical or critical), the response time of $\tau_o$ varies. For instance, when~$\tau_v$ executes its typical execution job, the response time of $\tau_o$ is $[9, 11]$ time units depending on whether $\tau_x$ executes typical or critical jobs in the $\lceil\frac{T_o}{T_v}\rceil$ interval. Likewise, when $\tau_v$ executes critical execution instances, the response time of $\tau_o$ becomes $[15, 17]$ time units depending on $\tau_x$'s execution. Thus, depending on the typical or critical job execution of higher-priority tasks, the observer task's response time changes, which, in this example, is in the range of $[9, 17]$ time units. This example suggests that even though the critical execution modes are non-deterministic, a preemptive priority scheduler may \textit{leak} arrival information about a high-priority task ($\tau_v$) to a low-priority one ($\tau_o$).

\begin{table}[!t]
\caption{Example Taskset.}
\centering
\begin{tabular}{|c|c|c|c|c|}
\hline
\multirow{2}{*}{\bfseries Task} & \multirow{2}{*}{\bfseries Priority} & \multirow{2}{*}{\bfseries Period} & \multicolumn{2}{c|}{\bfseries Execution Time} \\ \cline{4-5}
                      &                           &                          & \textbf{Typical}         & \textbf{Critical}         \\ \hline        
				\hline $\tau_v$ & High & 10 & 1 & 3 \\ 
				\hline $\tau_x$ & Medium & 15 & 2 & 3  \\
                \hline $\tau_o$ & Low & 30 & 2 & 2\\
				\hline
\end{tabular}	
\label{tab:example_data_1}
\end{table}



\subsection{Calculation of Response Times} \label{sec:rta}

The analysis we presented above is based on a simple window-based interference calculation that provides an upper bound of the response time. We now provide a tighter bound based on traditional real-time response time analysis~\cite{joseph_Response_time}. For a given $C_i$, the response time of $\tau_i$ is obtained by the following recurrence relation:
\begin{equation}
    R_i(k+1)=B_i+ C_i +\sum_{\tau_j \in hp(\tau_i)} \left\lceil \frac{R_i(k)}{T_j} \right\rceil C_j.
\label{eq:rta}
\end{equation}
The recurrence in Eq.~\eqref{eq:rta} starts with $R_i(0) = C_i$. The iteration terminates once the worst-case response time is obtained, \ie $R_i(k+1)=R_i(k)$ for some value of $k$ or when $R_i(k) > D_i$; in that case, the task is unschedulable.

We modify standard response time calculations for data-driven tasks. Let us define $C_i^{max} = \max(C_i^{typ}, C_i^{cri})$ and $C_i^{min} = \min(C_i^{typ}, C_i^{cri})$. The maximum (minimum) response time $R_i^{max}$ ($R_i^{min}$) obtained by replacing $C_i$ in Eq.~\eqref{eq:rta} with $C_i^{max}$ ($C_i^{min}$). Depending on which jobs (typical or crucial) the tasks execute, the actual response time of $\tau_i$ (denoted by $R_i^a$) varies as follows: $R_i^{min} \leq R_i^a \leq R_i^{max}$. 

Following noninterference rule~\cite{fixed_priority_sched} (see Section~\ref{sec:threat_model}), an adversary can calculate the ranges of victim's response times (\ie $[R_v^{min}, R_v^{max}]$) using offline calculations. However, due to the non-deterministic execution modes, the actual runtime response time $R_v^a$ cannot be inferred apriori. We now present an analysis technique that allows the observer task to infer the future arrivals of victim tasks by measuring the victim's own response times. The observer can extract the differences between its job arrival and completion (\ie $R_o^a$) by reading the values from the system clock.

\section{Inferring Typical and Critical Job Arrivals}\label{infer_pst}

Recall from earlier discussions (Section \ref{sec:infleak}) that our inference is based on measuring the response times of the observer. Hence, we want to derive the \textit{likelihood} (\ie conditional probability) of the runtime of $t$\textsuperscript{th} job of $\tau_o$ (denoted by $\hat{r}_o^{(t)}$) given a past record. Mathematically, this conditional probability is represented as follows: $\mathbb{P}(\hat{r}_o^{(t)} | \hat{r}_o^{(1)}, \hat{r}_o^{(2)}, \cdots, \hat{r}_o^{(t-1)})$. However, for any arbitrary invocation $t$, we need to keep track of all $t-1$ records from the beginning of system operations. Further, obtaining this conditional probability is challenging in practice, even for a smaller setup. 

Note that there are $2^{\lceil\frac{R_o^{max}}{T_v}\rceil}$ different victim's job invocation combinations (typical or critical) between the intervals of consecutive observer's jobs. 
As a concrete example, assume $T_v = 5$, $T_o = 15$, and let $R_o^{max}$ is some value $<5$ which makes the taskset schedulable. Suppose $\lceil\frac{R_o^{max}}{T_v}\rceil = 3$. Hence, there will be 3 job instances of the victim between two consecutive executions of the observer task. Let $\tau_v^{typ}$ and $\tau_v^cri$ denote the typical and critical jobs of the victim tasks, respectively. The possible combinations include  $\{(\tau_v^{typ}, \tau_v^{typ}, \tau_v^{typ}), (\tau_v^{typ}, \tau_v^{typ}, \tau_v^{cri}), \cdots ,(\tau_v^{cri}, \tau_v^{cri}, \tau_v^{cri})\}$, \ie a total of $2^3 = 8$ different invocation patterns which may result in non-unique response times for the observer. Now, including other high-priority tasks, there will be a total of $ \prod_{\tau_h \in hp(\tau_o)} 2^{\lceil\frac{R_o^{max}}{T_v}\rceil}$ many different invocation patterns. Based on offline response time calculations (Section \ref{sec:rta}), as we shall see in the paper, a probabilistic inference model in conjunction with a clustering-based classification technique (Section~\ref{pst_based_analysis}-Section~\ref{sec:clusting}), can predict response times and correlate with $\tau_v$'s job modes (\ie typical or critical) through runtime observations of $\tau_o$'s response times. 

Despite variations in runtime due to the execution of typical and critical mode jobs, a set of periodic task operations forms a \textit{pattern}. As presented next, it is possible to reduce the dimensionality of the conditional probability and \textit{learn} the response time behaviors from a smaller record due to the presence of patterns. For this, we use the probabilistic suffix tree (PST)~\cite{pst} and clustering technique~\cite{k_means} to isolate typical and critical instances. 

Suppose the observer extracts the following traces of response times from the system clock: $r_o^1, r_o^2, r_o^1, r_o^3, r_o^2, r_o^2$, where each $r_o^i,~i \in \{1,2,3\}$ represents a different response time observation. A
PST learns a set of subsequences of different lengths, \eg $\{ r_o^1 \}$, $ \{ r_o^3, r_o^3, r_o^2 \}$, each of which can be an indicator of
the execution mode of the victim's next job. The proposed PST-based inference enables us to calculate the probability of the victim's arrival and the job type (typical or critical) without having to look back at the entire history, that is, 
$\mathbb{P}(\hat{r}_o^{(t)} | \hat{r}_o^{(1)}, \cdots, \hat{r}_o^{(t-1)}) \sim \mathbb{P}(\hat{r}_o^{(t)} | \hat{r}_o^{(t-k)}, \cdots, \hat{r}_o^{(t-1)})$, for some predefined starting job $k$.

\paragraph*{Why Statistical Analysis?} One may wonder why we use statistical techniques like PST and clustering to analyze the victim's task arrivals. As we examine non-deterministic arrival patterns (and hence a non-uniform schedule, albeit constrained by real-time requirements), statistical techniques allow us to \textit{learn} and \textit{predict} from past observations. While we cannot ensure deterministic guarantees or provide bounds on successful inference, as we find from our evaluation, such a model can result in relatively fewer mispredictions. We intentionally avoid exploring any sophisticated learning techniques, such as neural networks, as our target domain is real-time, resource-constrained applications. Adding a complicated model just for inference may increase the observer task's response time and cause deadline violations, and ultimately, the inference will not remain stealthy. Besides, a majority of real-time operating systems do not support deep learning or other neural network libraries, which further limits practicality (in fact, making deep learning workloads real-time aware is an active area of research!). Instead, statistical analyses such as those introduced here are based on simple arithmetic computations, which most real-time and control tasks perform as a part of their routine operations.

\subsection{PST-based Prediction} \label{pst_based_analysis}



PST is a probabilistic model that represents sequences through a tree structure, where each node corresponds to a ``suffix'' of the sequence, and each node stores the probability distribution of ``symbols'' (\ie response times in our context) following that suffix. 
Constructing a PST involves extracting suffixes, calculating probabilities, and applying thresholds to ensure statistical significance and relevance. 


Let \( \mathcal{R} = \tilde{r}_1 \tilde{r}_2 \ldots \tilde{r}_\sigma \) be a sequence of response times of length \( \sigma \), where each symbol \( \tilde{r}_i \) belongs to a finite alphabet $\mathcal{A}$. In our setup, the members of the alphabet are in interval $[R_o^{min}, R_o^{max}]$. We build the PST by extracting suffixes from \( \mathcal{R} \) and storing them as nodes in a tree structure. Let us define a suffix \( \mathcal{S}_k = \tilde{r}_{k} \tilde{r}_{k+1} \ldots \tilde{r}_\sigma \) as the substring of \( \mathcal{R} \) starting at position \( k \) and extending to the end of the sequence. For each \( k \), we consider all possible suffixes of lengths \( 1 \) to \( L \), where \( L \) is the maximum allowable suffix length.

For each suffix \( \mathcal{S} \) of length \( |\mathcal{S}| \leq L \), we define the frequency counts as follows:
\begin{equation} \label{eq:ns}
N(\mathcal{S}) = \sum_{i=1}^{n-|\mathcal{S}|+1} \mathbb{I}(\mathcal{R}_{i:i+|\mathcal{S}|-1} = \mathcal{S}),    
\end{equation}
where \( \mathcal{R}_{i:i+|\mathcal{S}|-1} \) denotes the substring of \( \mathcal{R} \) starting at position \( i \) and ending at \( i+|\mathcal{S}|-1 \), and \( \mathbb{I}(\cdot) \) is the indicator function, which equals 1 if the condition is true and 0 otherwise. Let \( N(\mathcal{S} \tilde{r}_a) \) denote the count of occurrences where the suffix \( \mathcal{S} \) is immediately followed by the symbol $\tilde{r}_a \in \mathcal{A}$:
\begin{equation} \label{eq:nsa}
N(\mathcal{S}\tilde{r}_a) = \sum_{i=1}^{n-|\mathcal{S}|} \mathbb{I}(\mathcal{R}_{i:i+|\mathcal{S}|-1} = \mathcal{S} \ \text{and} \ \mathcal{R}_{i+|\mathcal{S}|} = \tilde{r}_a)    
\end{equation}
The conditional probability of observing the the response time \( \tilde{r}_a \) immediately after the suffix \( \mathcal{S} \) is given by:
\begin{equation} \label{eq:rs}
P(\tilde{r}_a | \mathcal{S}) = \frac{N(\mathcal{S} \tilde{r}_a)}{N(\mathcal{S})}. 
\end{equation}
This probability is stored at the node corresponding to the suffix \( \mathcal{S} \) in the tree.

For efficiency reasons, we do not require all suffixes to be included in the tree. For a suffix \( \mathcal{S} \) to be included in the tree, each probability \( P(\tilde{r}_a | \mathcal{S}) \) must satisfy the following condition:
$P(\tilde{r}_a | \mathcal{S}) > P_{min}$,
where \( P_{min} \) is a predefined minimum probability threshold. This criterion ensures that the probabilities are high enough to justify inclusion in the tree.
Finally, to update the probabilities, we sort the estimated probabilities \( P(\tilde{r}_a | \mathcal{S}) \) for each accepted node and then update the probability distributions for longer suffixes in an iterative manner. 

Once the PST is constructed, we can predict the future response times from the past sequence. Given a sequence of the observer's response times \( \mathcal{R}_o = r_o^1 r_o^2 \ldots r_o^m \), which we refer to as \textit{observation window}, the future response time calculation is based on probability values stored at the deepest matching node, \ie 
$\arg\max_{r_o^a \in \mathcal{A}} P(r_o^a | \mathcal{R}_o)=\arg\max_{r_o^a  \in \mathcal{A}} \frac{N( \mathcal{R}_o r_o^a )}{N(\mathcal{R}_o)}$.

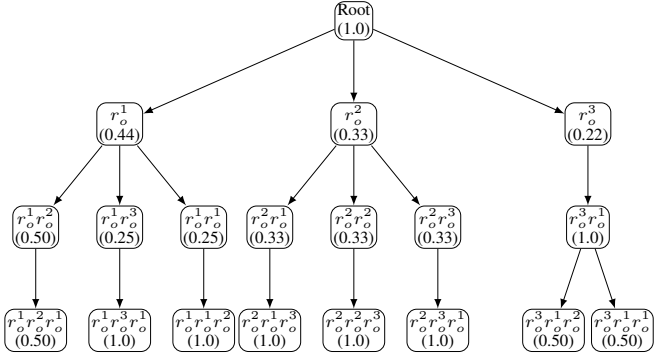
\begin{figure}[!ht]
\begin{center}
\resizebox{\columnwidth}{!}{
\begin{tikzpicture}[
  level distance=1.5cm, 
  level 1/.style={sibling distance=3.4cm}, 
  level 2/.style={sibling distance=1.22cm}, 
  level 3/.style={sibling distance=1cm},  
  edge from parent/.style={draw, -latex},
  every node/.style={
    draw, 
    rectangle, 
    rounded corners, 
    align=center, 
    font=\scriptsize, 
    inner sep=1pt 
  }
]

\node {Root \\ (1.0)}
  child {node {$r_o^1$ \\ (0.44)}
    child {node {$r_o^1r_o^2$ \\ (0.50)}
      child {node {$r_o^1r_o^2r_o^1$  \\ (0.50)}}
    }
    child {node {$r_o^1r_o^3$ \\ (0.25)}
      child {node {$r_o^1r_o^3r_o^1$ \\ (1.0)}}
    }
    child {node {$r_o^1r_o^1$ \\ (0.25)}
      child {node {$r_o^1r_o^1r_o^2$ \\ (1.0)}}
    }
  }
  child {node {$r_o^2$ \\ (0.33)}
    child {node {$r_o^2r_o^1$ \\ (0.33)}
      child {node {$r_o^2r_o^1r_o^3$ \\ (1.0) }}
    }
    child {node {$r_o^2r_o^2$ \\ (0.33)}
      child {node {$r_o^2r_o^2r_o^3$ \\ (1.0)}}
    }
    child {node {$r_o^2r_o^3$ \\ (0.33)}
      child {node {$r_o^2r_o^3r_o^1$ \\ (1.0)}}
    }
  }
  child {node {$r_o^3$ \\ (0.22)}
    child {node {$r_o^3r_o^1$ \\ (1.0)}
      child {node {$r_o^3r_o^1r_o^2$ \\ (0.50)}}
      child {node {$r_o^3r_o^1r_o^1$ \\ (0.50)}}
    }
  };

\end{tikzpicture}
}
\end{center}
\caption{A PST for a sequence of length 1200 generated
with the base pattern of $\mathcal{R} =r_o^1r_o^2r_o^1r_o^3r_o^1r_o^2r_o^2r_o^3r_o^1$. The maximum depth is set to 3.}
\label{fig:pst}
\end{figure}

\begin{example}
    Consider the victim's response time sequence $\mathcal{R} =r_o^1r_o^2r_o^1r_o^3r_o^1r_o^2r_o^2r_o^3r_o^1$ repeated to 1200 samples, where $r_o^i$ indicates the different response time of observer task $\tau_o$. Figure~\ref{fig:pst} depicts the resultant PST. Let us focus on the subsequence $r_o^1r_o^2$. To predict the next response time after $r_o^1r_o^2$, we examine the occurrences of subsequences following $r_o^1r_o^2$ in the extended sequence. From this sequence, the possible continuations after $r_o^1r_o^2$ include $r_o^1r_o^2r_o^1$, and $r_o^1r_o^2r_o^2$. Counting these occurrences, we find that $r_o^1r_o^2r_o^1$ appears 134 times, $r_o^1r_o^2r_o^2$ appears 133 times, totaling 267 instances. The probabilities for each continuation are then calculated as follows: $\mathbb{P}(r_o^1|r_o^1r_o^2r_o^1) = \frac{134}{267} \approx 0.50$, $\mathbb{P}(r_o^2 | r_o^1r_o^2r_o^2) = \frac{133}{267} \approx 0.49$. Thus, given the subsequence $r_o^1r_o^2$, the response time of the next invocation is equally likely to be $r_o^1$, $r_o^2$, each with a probability of approximately $0.50$ and $0.49$.
\end{example}

\subsection{Inference Algorithm}

\begin{algorithm}[!t]
\small
\caption{PST Construction and Inference}
\begin{algorithmic}[1]
\State \textbf{Input:} Sequence of response times $\mathcal{R} = \tilde{r}_1 \tilde{r}_2 \ldots \tilde{r}_\sigma$, suffix length \( L \), probability threshold \( P_{min} \), training time \(t_{train}\) Prediction duration \(t_{pred}\)
\State \textbf{Output:} Constructed PST and predicted response times 

\State \textbf{Initialize:} Set the root of the tree as an empty node.

\For{\( k = 1 \) to \( \sigma \)} 
    \State Extract suffix \( \mathcal{S}_k = \tilde{r}_{k} \tilde{r}_{k+1} \ldots r_\sigma \)
    \For{\( l = 1 \) to \( \min(L, \sigma - k + 1) \)}
        \State Define suffix \( \mathcal{S} \leftarrow \tilde{r}_{k} \ldots \tilde{r}_{k+l-1} \)
        \State Compute frequency count $N(\mathcal{S})$ using Eq.~\eqref{eq:ns}
        \For{each response time \( \tilde{r}_a \in \mathcal{A} \)}
            \State Compute count of \( \mathcal{S} \) followed by \( \tilde{r}_a \) using Eq.~\eqref{eq:nsa}
            
            \State Calculate  $P(\tilde{r}_a | \mathcal{S})$ using Eq.~\eqref{eq:rs}
            
            \If{\( P(\tilde{r}_a | S) > P_{min} \)}
                \State Add node \( \mathcal{S} \) with probability \( P(\tilde{r}_a | S) \)
            \EndIf
        \EndFor
    \EndFor
\EndFor

\State \textbf{Prediction:} 
\State Given a sequence \( \mathcal{R}_v \), identify the longest matching suffix \( \mathcal{S} \)

\For{\( time = t_{train} \) to \( t_{train}+t_{pred} \)} 
\State Predict the future arrival of $\tau_v$:
$\arg\max_{r_o^a \in \mathcal{A}} P(r_o^a | \mathcal{R}_o)$

\EndFor
\end{algorithmic}
\label{algo2}
\end{algorithm}


Algorithm~\ref{algo2} constructs a PST and predicts the future response times from a past sequence. The algorithm begins by initializing the PST with an empty root node. It iterates over each position \( k \) in the sequence \( \mathcal{R}  \), extracting suffixes \( \mathcal{S} \) starting from each position (Line 5). For each suffix \( S \) of length 1 up to the maximum allowable length \( L \), the frequency count \( N(\mathcal{S}) \) of the suffix within the sequence is computed (Line 8). For each possible response time candidate \( \tilde{r}_a \) from the alphabet \( \mathcal{A} \), the count \( N(\mathcal{S}\tilde{r}_a) \) of occurrences where \( S \) is immediately followed by \( \tilde{r}_a \) is determined (Line 10). Then we calculate the conditional probability \( P(\tilde{r}_a | \mathcal{S}) \) of observing \( \tilde{r}_a \) after \( \mathcal{S} \) (Line 11). If this probability exceeds a predefined threshold \( P_{min} \), the suffix \( \mathcal{S} \) is added as a node in the tree with the corresponding probability (Line 13). Finally, the PST uses the longest matching suffix of a given sequence $\mathcal{R}_o = r_o^1 r_o^2 \ldots r_o^m$ to predict the next symbol based on the highest stored conditional probability (Line 21).

\subsection{Classify Victim's Instances} \label{sec:clusting}

Recall from our earlier discussion in the beginning of Section~\ref{infer_pst}, due to typical and critical invocation patterns of the higher-priority tasks that the observer in an interval of size $T_v$, there are $ \prod_{\tau_h \in hp(\tau_o)} 2^{\lceil\frac{R_o^{max}}{T_v}\rceil}$ different response times for the observer task. As the observer task may observe several response times due to complex arrival patterns of the higher priority tasks, we need to correlate its response times with the actual execution mode of the victim (especially the victim's critical jobs). The PST-based inference presented above helps us to predict the response times of the $\tau_o$ based on past observations. Based on this prediction, we adopt a clustering technique to classify which response times of $\tau_o$ resulted due to the critical (resp. typical) invocation of the victim task. Our reasoning is that, as critical execution modes have longer execution durations in general, when many higher-priority tasks execute their critical jobs, the response times of the observer task will be higher. Hence, it is possible to cluster the response times of $\tau_o$ into two buckets and use this information to filter out critical (and typical) jobs of the victim task.

We use K-means\footnote{K-means is an unsupervised clustering algorithm used to partition a dataset into \(k\) distinct, non-overlapping groups or clusters. The algorithm minimizes the sum of squared distances between the data points and their corresponding cluster centroids. The goal of K-means is to minimize the within-cluster sum of squares (WCSS)~\cite{wcss}. Mathematically, this is expressed as: $J = \sum_{i=1}^{k} \sum_{x \in \mathcal{C}_i} \| x - \mu_i \|^2$, where \(k\) is the number of clusters, \(\mathcal{C}_i\) represents the set of points belonging to cluster \(i\), \(x\) is a data point in cluster \(\mathcal{C}_i\), \(\mu_i\) is the centroid of cluster \(\mathcal{C}_i\), \(\| x - \mu_i \|\) represents the Euclidean distance between a point \(x\) and the centroid \(\mu_i\). 
} clustering~\cite{k_means} to identify a threshold to separate the observer's response times (\ie which of those caused by the victim's typical jobs and which are caused by the victim's critical jobs) considering arbitrary execution patterns of other higher-priority tasks between victim and observer (\ie $\forall \tau_h \in hp(\tau_o) - \{\tau_v\}$). Our intuition is that for a given victim's typical job and irrespective of arbitrary typical/critical instances of other higher-priority tasks $\tau_h \in hp(\tau_o) - \{\tau_v\}$, we put the response times of $\tau_o$ to one cluster. Likewise, we do the same for the victim's critical arrivals and isolate the observer's response times to another cluster. At runtime, depending on which cluster the predicted response time maps (\ie obtained from Algorithm~\ref{algo2}), $\tau_o$ can correlate it with the actual typical or critical invocation of the victim task.

\begin{table}[!t]
\caption{Example Taskset to Illustrate Clustering.}		
\centering
\begin{tabular}{|c|c|c|c|c|}
\hline
\multirow{2}{*}{\bfseries Task} & \multirow{2}{*}{\bfseries Priority} & \multirow{2}{*}{\bfseries Period} & \multicolumn{2}{c|}{\bfseries Execution Time} \\ \cline{4-5}
                      &                           &                          & \textbf{Typical}         & \textbf{Critical}         \\ \hline
                      \hline $\tau_v$ & 1 (High) & 30 & 2 & 6 \\ 
				\hline $\tau_1$ & 3 & 80 & 4 & 6  \\
                \hline $\tau_2$ & 2 & 70 & 5 & 8  \\
                \hline $\tau_3$ & 4 & 90 & 15 & 15  \\
                \hline $\tau_o$ & 5 (Low) & 100 & 12 & 12\\
				\hline
\end{tabular}

\label{tab:example_data_2}
\end{table}


\begin{figure}[!ht]
	\centering
	\includegraphics[scale=0.45]{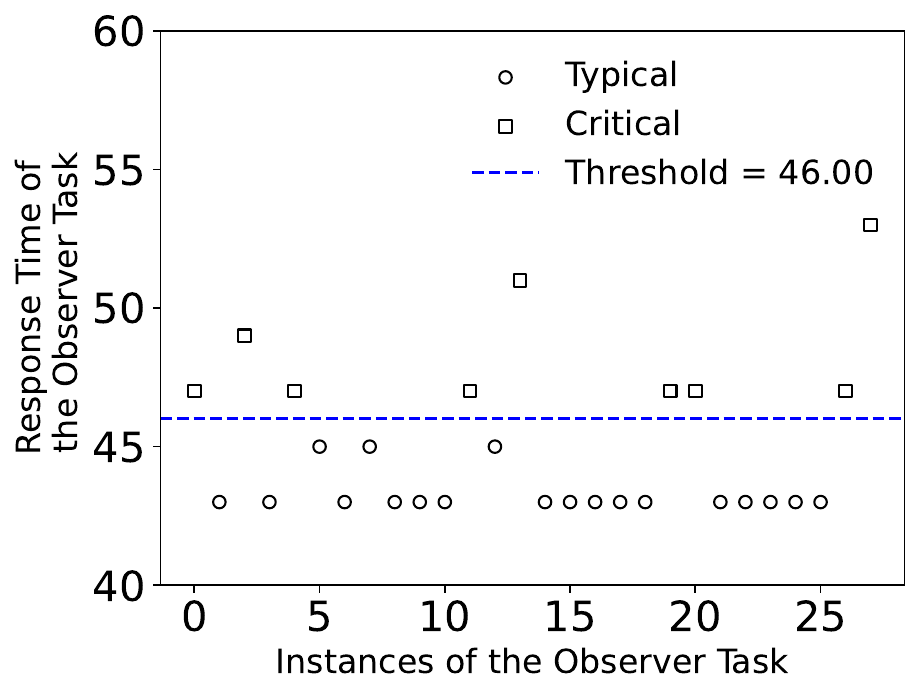}
	\caption{Using K-means to isolate $\tau_o$'s response times into two clusters. Instances of an observer task with response times above the threshold are caused due to critical invocation of the victim (square), while those that are below result in due to typical invocation of the victim (circle).}
	\label{fig:k-means clustering}
\end{figure}

To illustrate the idea, consider the taskset presented in Table~\ref{tab:example_data_2}. Here, the victim task's typical execution time is 2, and the critical execution time is 6. We train the PST for $\hat{H}$ hyperperiods (where the numbers vary as follows: $\hat{H} \in [2,5,10,20,30,50,70,90,110]$) and calculate the response time of \(\tau_v\). In this setup, the measured response times of the observer task are $43, 47, 51, 45, 49$. We apply K-means to partition the observer's response times into two clusters and use it as a cutoff to distinguish the victim's arrival modes. For instance, as Fig.~\ref{fig:k-means clustering} shows, if a predicted response time of $\tau_o$ exceeds this threshold (46 in this case, the cutoff point for the clusters), it is due to critical invocation of $\tau_v$; otherwise, it is due to typical invocation of $\tau_v$.

\subsection{Timing Considerations}


\begin{figure}[htbp]
\centering
\includegraphics[scale=0.7]{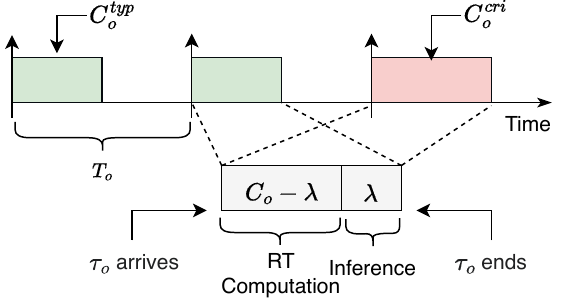}
\caption{Execution of $\tau_o$ has $C_o-\lambda$ unit of time for computation and $\lambda$ time for inference.}
\label{fig:low_response_fraction}
\end{figure}

For correct system operation (and to remain stealthy), $\tau_o$ should not run more than its worst-case execution time, $C_o$. Hence, it saves some budget to perform PST-based inference. We define a parameter, $\lambda$, as shown in Fig.~\ref{fig:low_response_fraction}, whose value is set by the observer to limit the running time of the inference function in each period. This inference duration $\lambda$, is an integer in the range $0 < \lambda < C_o$, as shown in Fig.~\ref{fig:low_response_fraction}. Note that the PST and clustering threshold can be built offline from public knowledge (recall: we follow the non-interference model) and embedded in the task logic (\ie part of the code binary). We empirically measured the value of $\lambda$ in a Raspberry Pi platform~\cite{Raspberrypi} and found that the inference operations do not have significant overheads (about 50  milliseconds; see additional discussions in Section~\ref{sec:overhead_pi}).

\section{Evaluation}\label{sec:eval}

We evaluate our framework on two fronts: \ca evaluation with synthetic tasks to check the feasibility of the inference (Section~\ref{simulation}), and \cb case-study using a UAV autopilot system (Section~\ref{sec:arduplilot}). We also measure the overheads of the model (Section~\ref{sec:overhead_pi}). To this end, we carried out experiments to demonstrate real-world applicability using two cyber-physical platforms: a robotic arm and a surveillance system (Section~\ref{sec:cps_demo}).

\subsection{Evaluation with Synthetic Tasks}\label{simulation}


\begin{table}[!t]
\caption{Simulation Parameters.\vspace{-0.5\baselineskip}}
\setlength{\tabcolsep}{2.8pt}
\begin{center}
\begin{tabular}{|l||l| } 
 \hline
\textbf{Parameters} & \textbf{Value}\\ 
\hline
\hline Utilization, $U$  & 0\%-90\% \\
\hline Period $T$ & $[100,900]$~ms \\
\hline Hyperperiod  & $4500$~ms \\
\hline Number of tasks, $n$ & [7, 20] \\
\hline Number of taskset for each utilization, $N_u$  & $100$ \\
\hline
\end{tabular}
\end{center}
\label{tab:paramtable}
\end{table}

\subsubsection{Taskset and Parameters}

 
 Our first set of experiments analyze information leakage using synthetically generated tasksets with parameters similar to those used in prior work~\cite{isroc_mfb,deeptrust_babar}. We vary the system utilization from 0\% to 90\%. For each system utilization $u$ in the range $[0, 10, \cdots, 90]\%$, we generated $N_u = 100$ tasksets, each taskset containing $[7, 20]$ tasks. Task periods were randomly selected from $[100, 900]$ ms. The taskset was generated following uniform distribution using the UUniFast algorithm~\cite{uunifast}. The tasks follow the rate monotonic (RM)~\cite{rate_monotonic} scheduling policy. The tasksets were generated for a fixed hyperperiod of 4500 ms. We did so to ensure fair comparisons among different tasksets for different utilizations. Given the utilization, task period, and fixed hyperperiod, the worst-case execution time \(C_i = \max(C_i^{typ}, C_i^{cri})\) was computed. We assumed $C_i^{typ} \leq  C_i^{cri}$ and assigned $C_i^{typ} = 0.7 C_i^{cri}$. We set the critical arrival rate of tasks at $10\%$ to $30\%$ and varied it as an experiment parameter. Given the arrival rate, the jobs were marked as typical or critical following a uniform distribution. We set $P_{min} = 0.001$ in the experiments. Table~\ref{tab:paramtable} lists key simulation parameters.

 

\subsubsection{Reference Scheme: Random Selection}  \label{sec:random_selection}

As we want to analyze information leakage and predict (critical) job arrivals in a relatively non-deterministic setup, we use a ``random selection'' scheme as a baseline. At the high level, from the attacker's point of view, they need to consider whether a given arrival of a victim job is typical or critical. We test this prediction problem with a simple intuitive idea. Specifically, our baseline scheme \textit{randomly makes a decision} by mimicking this as a ``coin toss'' scenario, which follows an uniform distribution. For a target victim job $\tau_v^j$, the attacker flips a coin, and if it is head, then $\tau_v^j$ is assumed to be a typical job; otherwise, $\tau_v^j$ is considered as critical. While conceptually, the baseline scheme has a 50\% chance of being correct (\ie choose from one of two possible outcomes---typical or critical), as our results show (see Section~\ref{sec:sim_exp}), this may not be a good strategy to identify target arrivals (for instance, the critical jobs). This is due to the fact that \ca typical and critical arrivals are non-deterministic and \cb possibilities of critical arrivals are lower than typical, which results in non-uniform distribution of arrival patterns.

\subsubsection{Evaluation Metric and Test Cases}

We use two metrics to assess information leakage and analyze the goodness of prediction for both the proposed scheme and random selection.

\vspace{0.5\baselineskip}

\noindent
$\blacktriangleright$~~\textbf{Inference Precision.}\quad  Our first metric named \textit{inference precision}, which quantifies the effectiveness of the observer in successfully predicting the typical or critical mode of the victim task. We define inference precision as follows: 
\[
IP = \frac{\text{Successful Number of Predictions}}
{\text{Total Observed Jobs}}.
\]

\noindent
$\blacktriangleright$~~\textbf{False Positive.}\quad We also measure \textit{false positive} percentages. An inference is considered a false positive if a victim's arrival is predicted as critical, but actually, the job is typical. False positives are crucial in our study as often the correct identification of critical instances can be used as a precursor to launch a successful attack~\cite{scheduleak}. Higher false positives imply that a scheme is unable to gauge the typical arrivals correctly for most cases and, hence, may miss target attack points. 


\subsubsection{Experiment Setup}

 For our synthetically generated tasksets, we conducted three sets of experiments, as summarized below.

 \begin{itemize}
     \item \textbf{Experiment 1:} This experiment was designed to observe inference precision (\ie correctly determine typical and critical instances of $\tau_v$) for utilizations ranging from 0\% to 90\% and for a training duration of $\hat{H} = 50$ hyperperiods and runtime history length of past $|\mathcal{R}_o| = 20$ response times observations, with a critical arrival rate of victim set to \( 30\% \).

     \item \textbf{Experiment 2:} In this experiment, we observe false positives using a setup identical to that of Experiment 1.

     \item \textbf{Experiment 3:} The third experiment is to study statistical model internals, in particular the impact on the length of the past observations ($|\mathcal{R}_o|$). We observe the inference precision on a moderately loaded system (at 60\% utilization) and varying arrival rate of critical arrival rate of $\tau_v$ from \(10\%~\text{to}~30\%\)). In this setup, we tested with a training duration of 50 hyperperiods as before but varied the history length (\eg  $|\mathcal{R}_o| \in [2, 3, 4, 10, 20, 30, 40, 50]$).
     
 \end{itemize}

The workflow of our experiments was as follows. For a given trace of response time observations, we randomly selected a victim job (say $\tau_v^j$) and tried to infer whether $\tau_v^j$ was typical or critical. For our proposed statistical model (\eg PST and clustering), $\tau_o$ uses its learned PST and clustering cutoff threshold (offline knowledge). Based on this knowledge, at runtime (\ie while making the inference), $\tau_o$ observes its own past $|\mathcal{R}_o|$ observations to make a prediction on $\tau_v^j$. In the random selection scheme, we used a coin-toss test, as described earlier (Section~\ref{sec:random_selection}), to determine if  $\tau_v^j$  is typical or critical based on whether the flipped coin lands on heads or tails.



\begin{figure}
    \centering
    \begin{subfigure}{0.9\linewidth}
        \centering
        \includegraphics[scale=0.3]{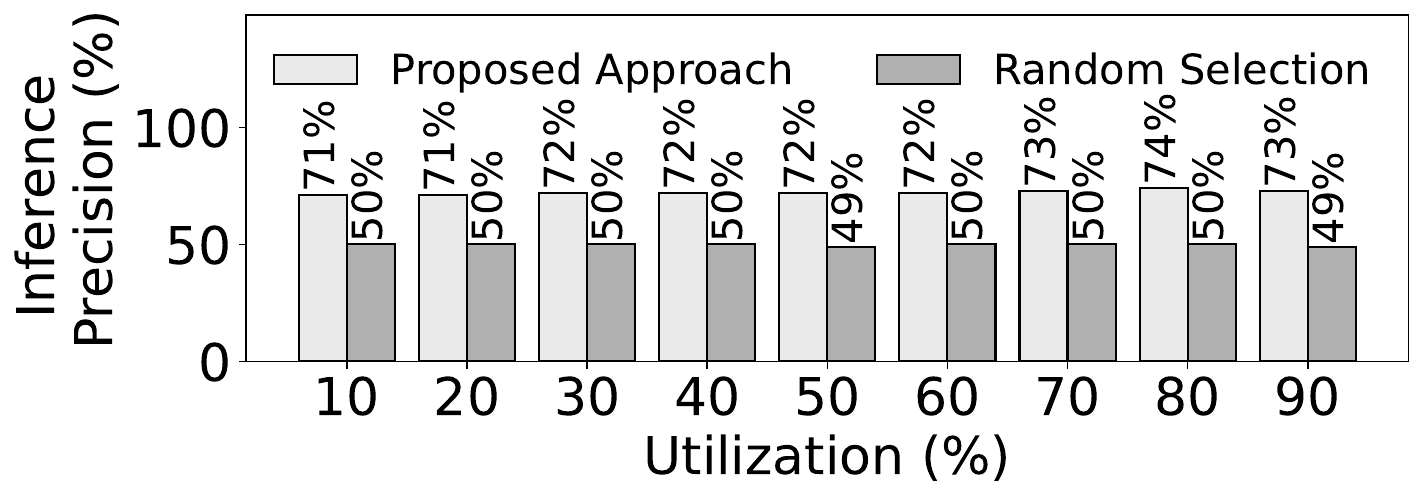}
        \caption{Case I: $10\%$ of $\tau_v$'s jobs are critical.}
        \label{fig:scheme1_ip_cric_10}
    \end{subfigure}
    \hfill
    \begin{subfigure}{0.9\linewidth}
        \centering
        \includegraphics[scale=0.3]{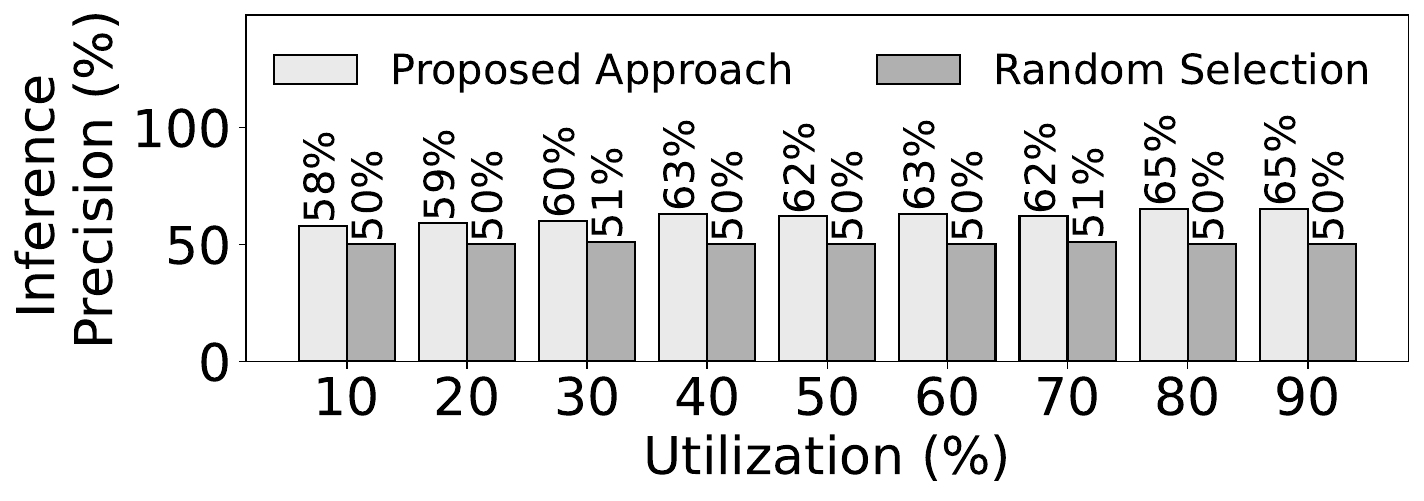}
        \caption{Case II: $20\%$ of $\tau_v$'s jobs are critical.}
        \label{fig:scheme1_ip_cric_20}
    \end{subfigure}
    \hfill
    \begin{subfigure}{0.9\linewidth}
        \centering
        \includegraphics[scale=0.3]{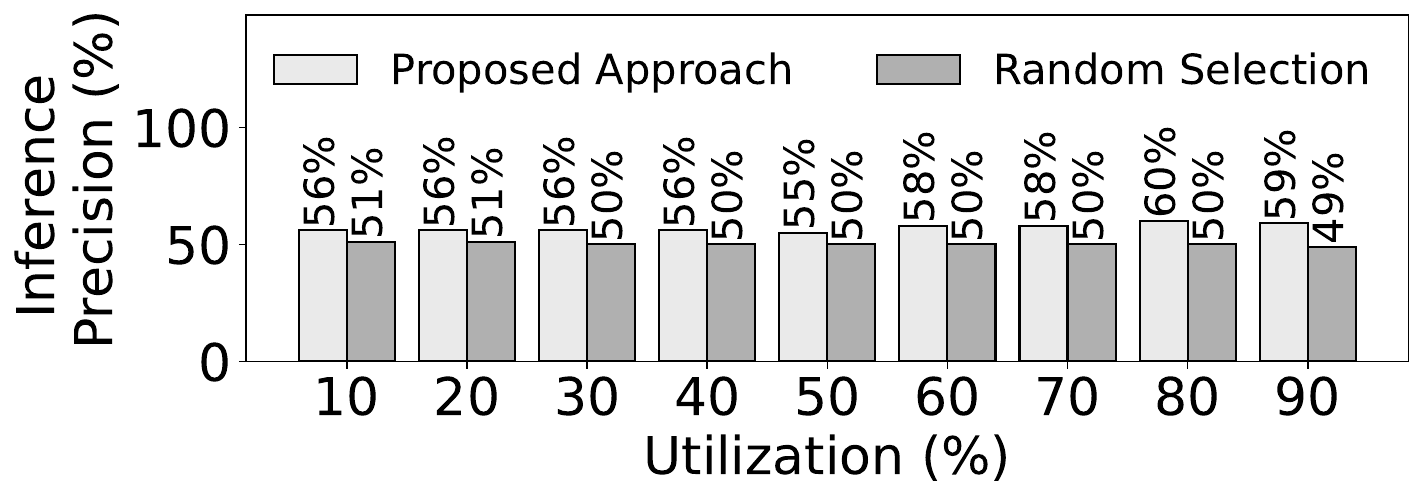}
        \caption{Case III: $30\%$ of $\tau_v$'s jobs are critical.}
        \label{fig:scheme1_ip_cric_30}
    \end{subfigure}
    \hfill
    \caption{Utilization vs precision. The statistical analysis introduced in this work outperforms the random selection. While system utilization does not impact the inference, when critical jobs arrive at a relatively higher rate, albeit in arbitrary instances, it becomes harder to predict them.}
    
    \label{fig:exp_set_1}
\end{figure}

\subsubsection{Results}  \label{sec:sim_exp}

In our first experiment (Fig.~\ref{fig:exp_set_1}),  we analyze inference goodness using the precision ratio metric for different system utilization. In this setup, we trained our statistical model for 50 hyperperiods to build the PST and get clustering thresholds. The likelihood of $\tau_v$'s critical workload arrival was varied and tested for three cases: \ca $10\%$,~\ie the majority of jobs ($90\%$) are executed on typical mode (Fig.~\ref{fig:scheme1_ip_cric_10}), \cb $20\%$ (Fig.~\ref{fig:scheme1_ip_cric_20}), and \cc $30\%$ (Fig.~\ref{fig:scheme1_ip_cric_30}). Note that as critical events are not so common, \ie  often invoked to tackle crucial system conditions (and hence are of interest to an adversary), we limit the critical job arrivals to $30\%$. We show the inference precision for our approach and the baseline (random selection). When typical jobs are dominating (Case~I), the schedule is somewhat predictable, and we get more than $70\%$ successful inference. As the results show, there is not much impact on system utilization. Although the results vary slightly due to random invocations of typical and critical jobs, they are all within two standard deviation ranges (\ie $\pm 0.91$). For the other two cases (Case~II and Case~III), the precision drops as higher critical job arrival rates of the victim result in more ``noisy'' observations from the observer's point of view. Hence, the model is unable to predict some jobs. Still, our proposed statistical analysis achieves more than $55\%$ of correct predictions and outperforms the random selection scheme. 

Note that the random selection outputs $~50\%$ correct inference for most cases. This result is not surprising, as for a large number of traces, statistically, a random coin flip has an equal probability of being heads and tails. Intuitively, this implies that the random selection can guess the typical/critical arrivals half of the time. However, as we will show in our next experiment, since the typical and critical job arrival intervals are not periodic (\ie they form non-uniform distribution), any random selection ends up being a wild guess and results in more false positives, particularly when identifying critical instances.

\begin{center}
\fbox{\begin{minipage}{0.97\linewidth}
\itshape
Statistical models like the one we introduced here, which use past information,  can achieve better inference precision compared to wildly guessing the victim's typical arrivals. 
\end{minipage}}
\end{center}

\begin{figure}[!t]
    \centering
    \begin{subfigure}{0.90\linewidth}
        \centering
        \includegraphics[scale=0.3]{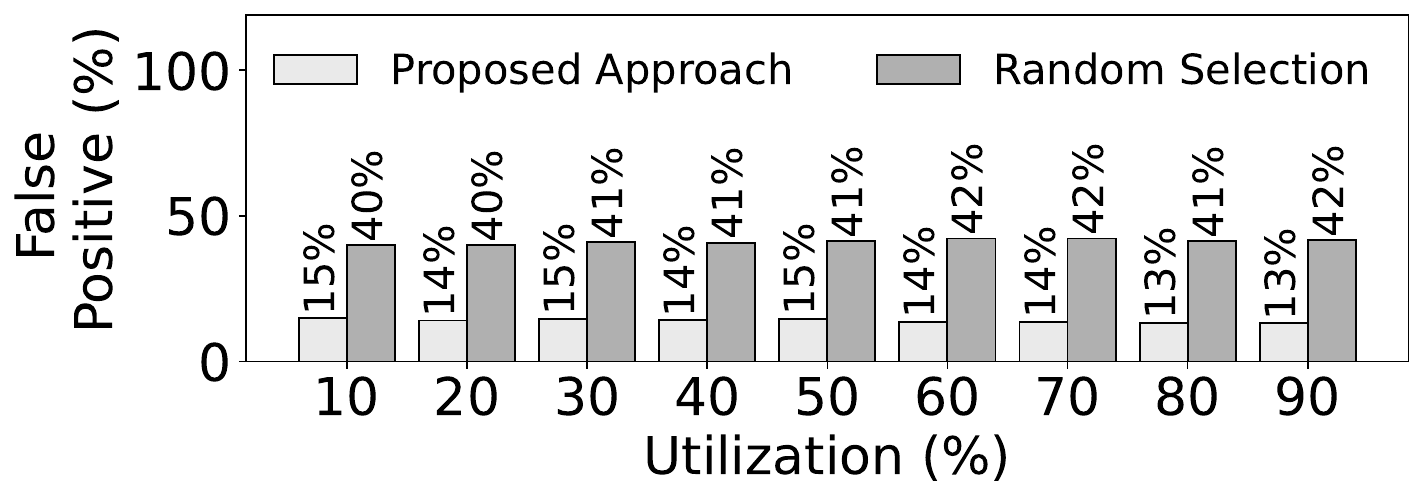}
        \caption{Case I: $10\%$ of $\tau_v$'s jobs are critical.}
        \label{fig:scheme1_fp_typ_10}
    \end{subfigure}
    \hfill
    \begin{subfigure}{0.90\linewidth}
        \centering
        \includegraphics[scale=0.3]{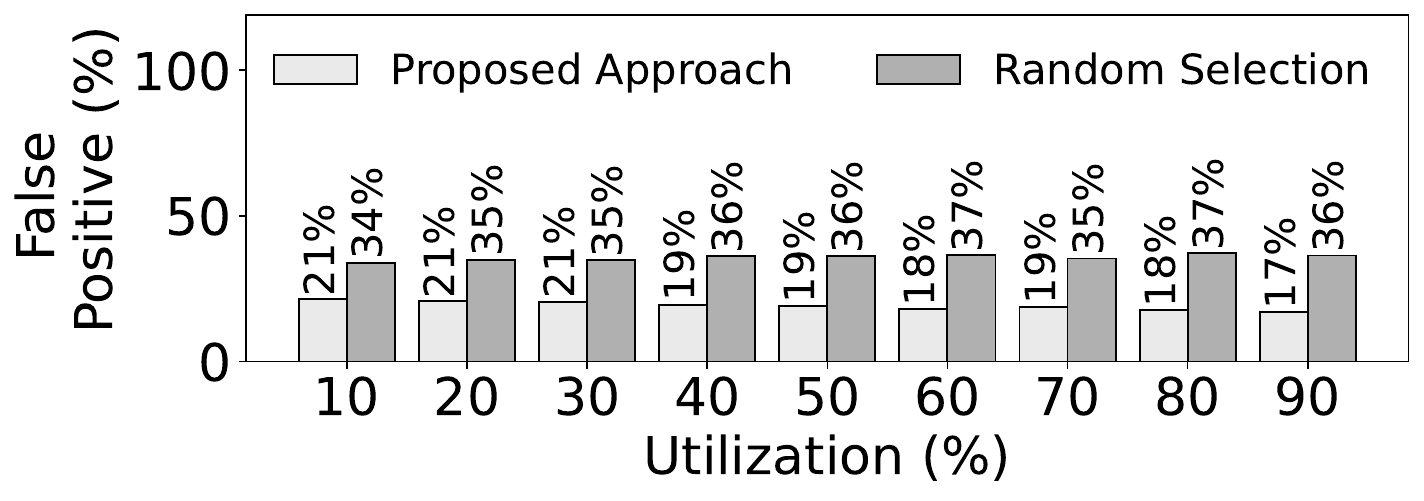}
        \caption{Case II: $20\%$ of $\tau_v$'s jobs are critical.}
        \label{fig:scheme1_fp_cric_20}
    \end{subfigure}
    \hfill
    \begin{subfigure}{0.90\linewidth}
        \centering
        \includegraphics[scale=0.3]{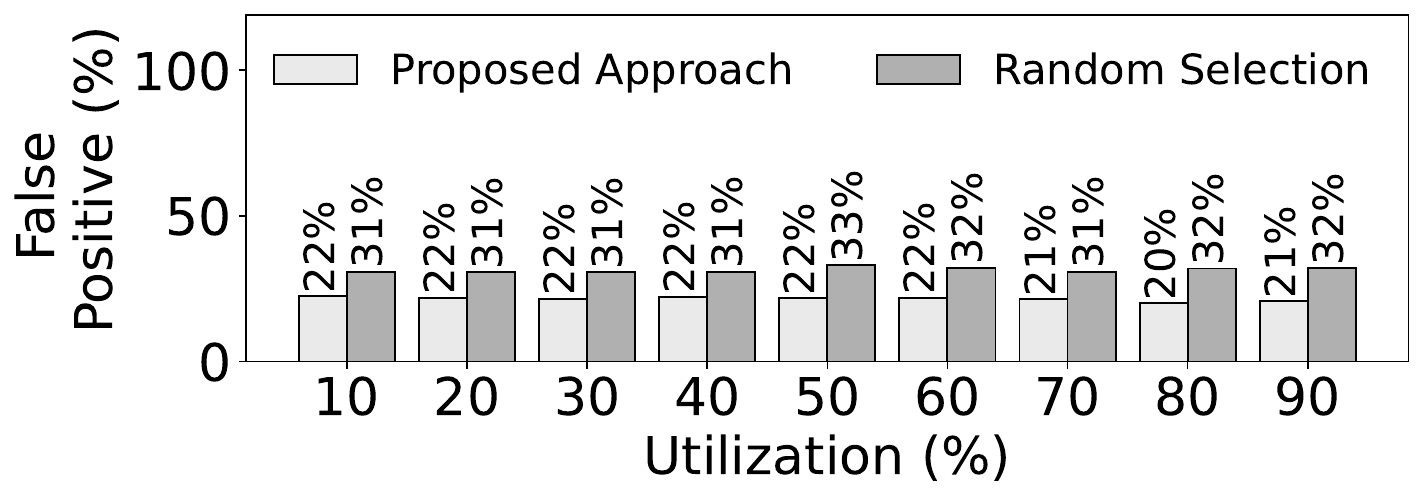}
        \caption{Case III: $30\%$ of $\tau_v$'s jobs are critical.}
        \label{fig:scheme1_fp_typ_30}
    \end{subfigure}
    \hfill
    \caption{Utilization vs false positive. The statistical test uses the correlation between interference caused due to victim's jobs and corresponding response times of the observer's jobs, which result in fewer false positives. In contrast, random guess failed to correctly predict, especially when critical events are rare (e.g., for Case I). }
    \label{fig:exp_set_2}
\end{figure}

In the next experiment (see Fig.~\ref{fig:exp_set_2}), we use an identical setup to the previous one but report the false positives. Recall that false positives numbers show the fraction of instances $\tau_o$ mispredict (\ie incorrectly classify a typical job as critical). As the figures show, the false positives for our inference techniques are significantly low (less than $25\%$). For less frequent critical arrivals (\eg Case~I and Case~II), the false positives are even lower ($15\%$-$20\%$ range). As was in the previous case, when there are more critical jobs, say for 30:70 of critical and typical splits (albeit in an irregular manner), this creates more noise in the receiver's observations. Due to this, the model fails to predict the critical invocations correctly (resulting in higher false positives). An interesting observation is the increase in false positives for the random selection scheme. As the random selection does not use any past observations or factor any correlations with the observer's response time to the victim's jobs, this ends up as a random guess, which may only occasionally be correct. This is more apparent for Case~I, when there are less frequent critical jobs and random selection results in $\sim\!40\%$ false positives. 

\begin{center}
\fbox{\begin{minipage}{0.97\linewidth}
\itshape
False positives are a crucial indication to assess the successful prediction of victim task's critical jobs. Our proposed statistical analysis results in low false positives (less than 25\%). If the attacker were to guess a critical job randomly, the false positive rates could be as high as 40\%, which may limit the chances of launching a targeted attack at the desired times (bad for adversary). 
\end{minipage}}
\end{center}

\begin{figure}[!t]
    \centering
    \begin{subfigure}{0.49\linewidth}
        \centering
        \includegraphics[width=\linewidth]{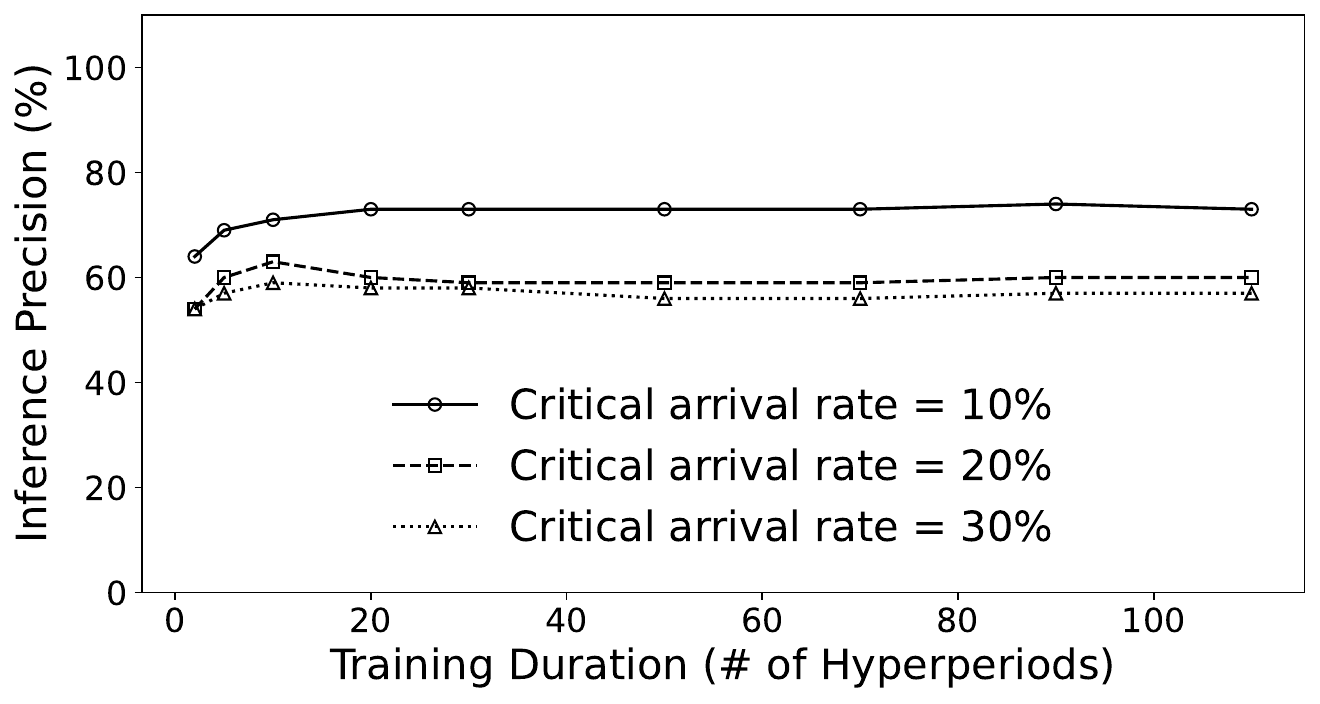}
        \caption{Training duration vs inference precision.}
        \label{fig:exp2_ip}
    \end{subfigure}
    \begin{subfigure}{0.49\linewidth}
        \centering
        \includegraphics[width=\linewidth]{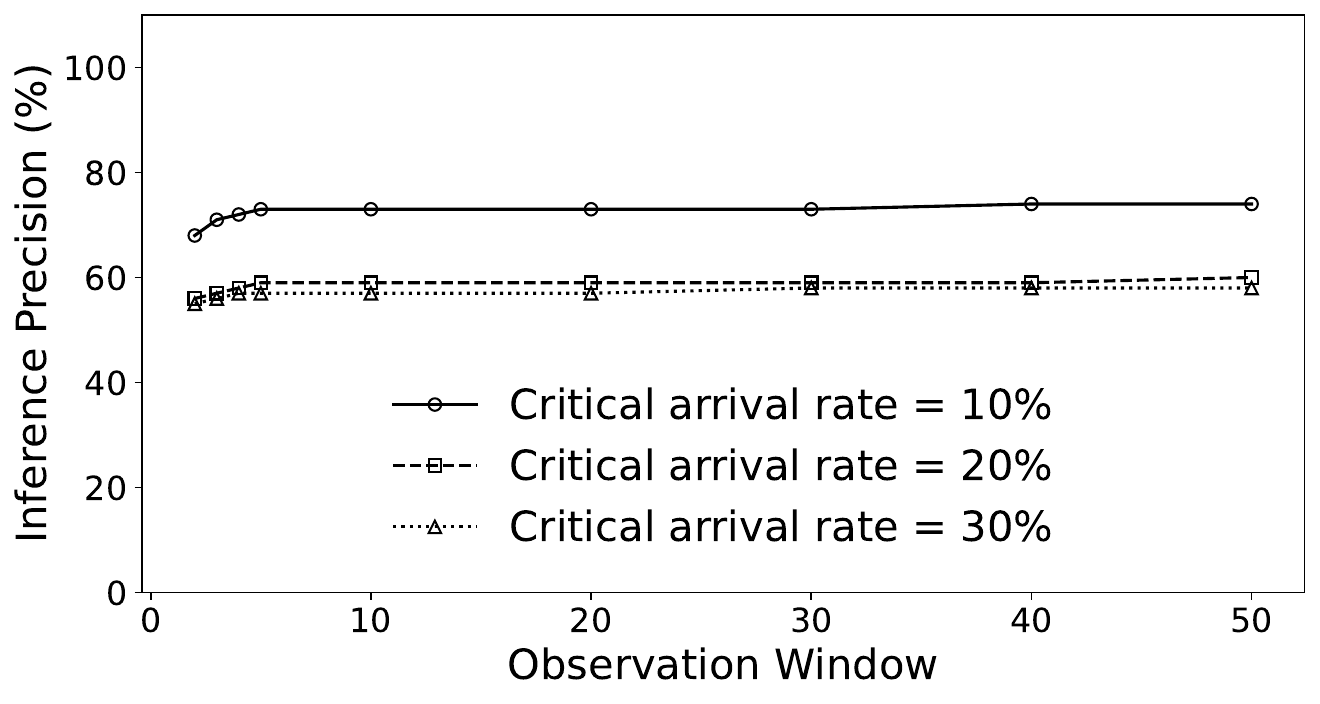}
        \caption{Observation window vs inference precision.}
        \label{fig:exp4_ip}
    \end{subfigure}
    \caption{Inference precision for varying training duration (left) and observation window size (right). Increasing training duration or having more past observations does not help the observer task to improve its inference performance.}
    \label{fig:ip_combined}
\end{figure}

\begin{figure}[!t]
    \centering
     \begin{subfigure}{0.49\linewidth}
        \centering
        \includegraphics[width=\linewidth]{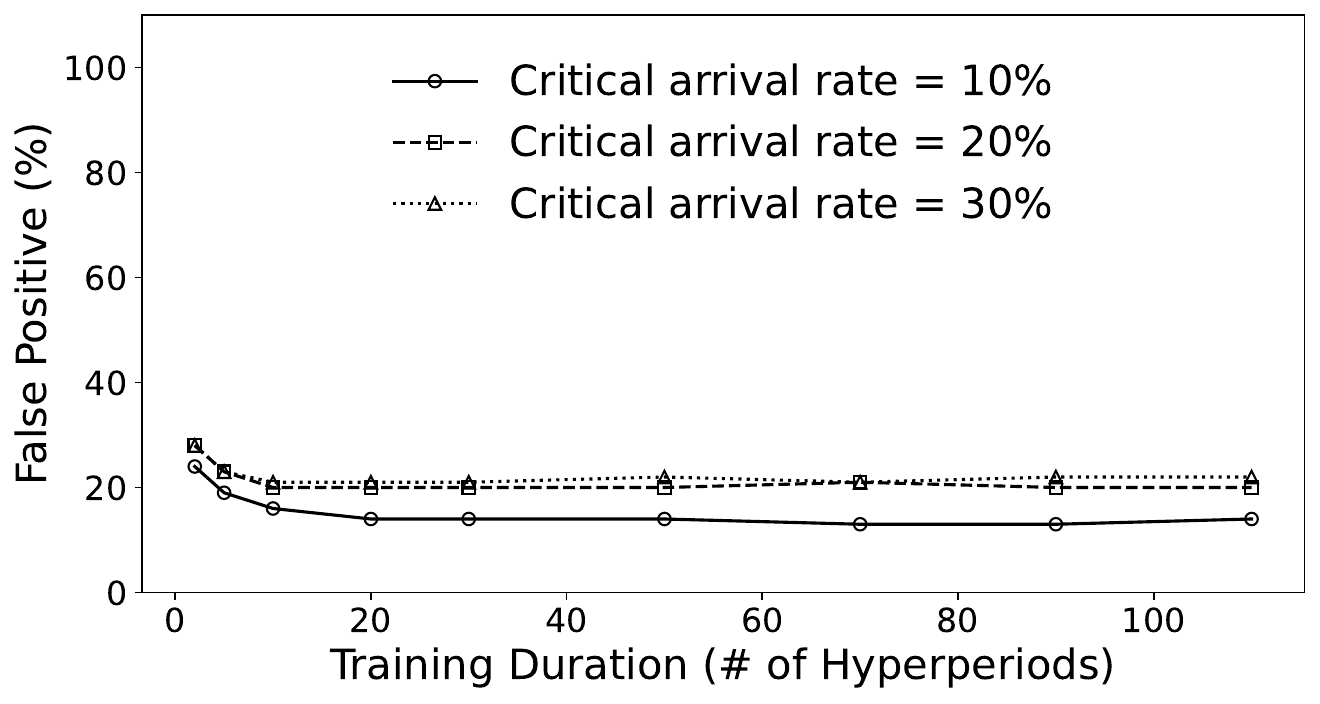}
        \caption{Training duration vs false positive.}
        \label{fig:exp2_fp}
    \end{subfigure}
    \hfill
    \begin{subfigure}{0.49\linewidth}
        \centering
        \includegraphics[width=\linewidth]{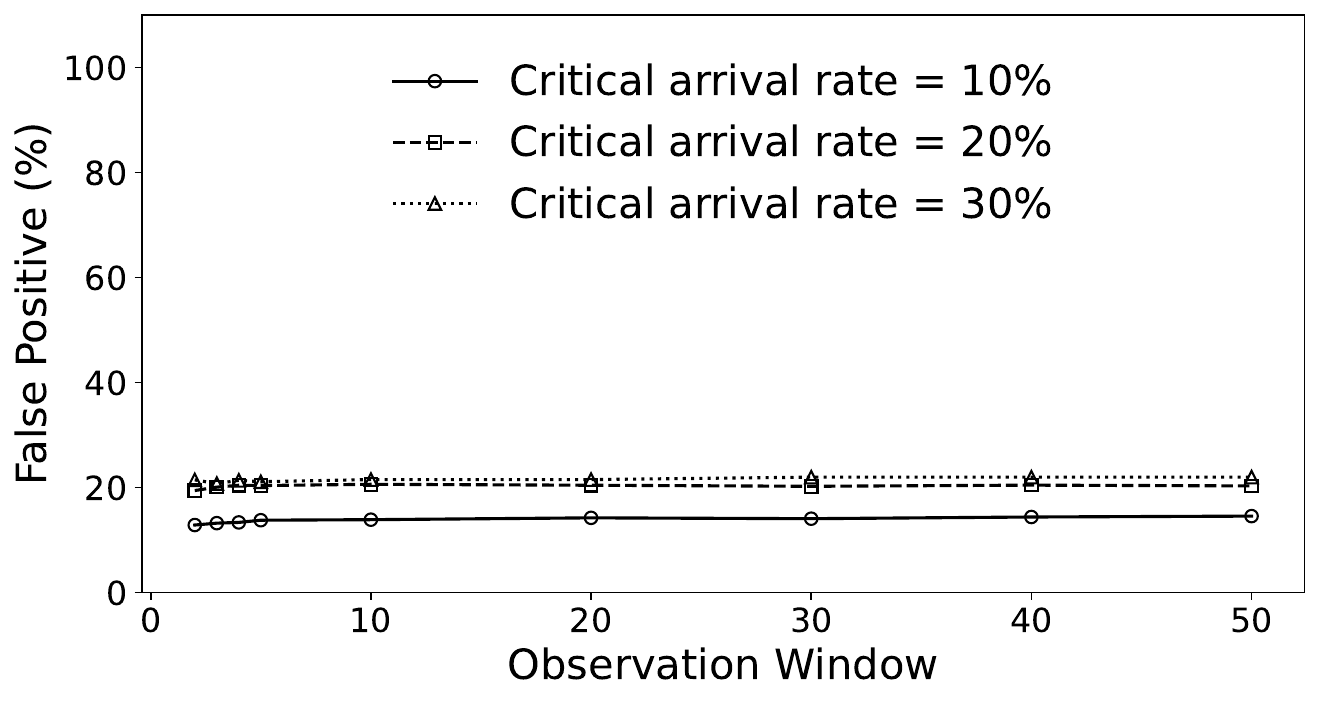}
        \caption{Observation window vs false positive.}
        \label{fig:exp4_fp}
    \end{subfigure}
    \caption{False positive rates for varying training duration (left) and observation window size (right). The findings are similar to that of Fig.~\ref{fig:ip_combined}---the false positive rates remain flat with increasing training duration and observation window.}
    \label{fig:fp_combined}
\end{figure}

In our third experiment, we study whether the training duration and the length of past observations (\ie observation window) impact inference. In Fig.~\ref{fig:exp2_ip}, we varied the training duration from 2 to 110 hyperperiods. In Fig.~\ref{fig:exp4_ip}, we built the PST and formed the clusters by training the model for 50 hyperperiods as before. For both setups, we show the inference precision for varying the observation window size ($|\mathcal{R}_o|$) for the three cases introduced earlier (\eg varying $\tau_v$'s critical jobs from $10\%$ to $30\%$). We considered a moderate system load ($60\%$ utilization). 
As Fig.~\ref{fig:ip_combined} shows, for infrequent $\tau_v$'s critical arrival (\eg Case~I, where $10\%$ of $\tau_v$'s jobs are critical), higher determinism in the schedule results in better inference precision. These findings are consistent with what we observe in Fig.~\ref{fig:exp_set_1}. The results also suggest that a longer training duration or a large number of past response time observations do not significantly improve inference precision or reduce false positive percentages. From $\tau_o$'s view, it is sufficient to observe only a few of its own past response times (\eg 10) to make inference decisions. We repeated the experiments to calculate the false positive rates (see Fig.~\ref{fig:fp_combined}), and our findings are similar.

While training is done offline and the inference logic is loaded as a part of the binary (\eg the $\lambda$ component in Fig.~\ref{fig:low_response_fraction}), the inference is conducted at runtime. Hence, this observation is beneficial from the attacker's perspective, as keeping and parsing large traces may increase the memory/timing overheads for $\tau_o$'s binary, and it may not remain stealthy.


\begin{center}
\fbox{\begin{minipage}{0.97\linewidth}
\itshape
The inference precision does not significantly improve even if we have large past observations. An observer task can make inference decisions by (offline) training of the schedule for 30 hyperperiods. During the runtime inference phase, tracking only a few (\eg 10) of the observer's past response times is sufficient to make the prediction. 
\end{minipage}}
\end{center}


\subsection{Case Study using a UAV Autopilot Taskset} \label{sec:arduplilot}

\begin{figure}[!t]
    \centering
    \begin{subfigure}{0.49\linewidth}
        \centering
        \includegraphics[width=\linewidth]{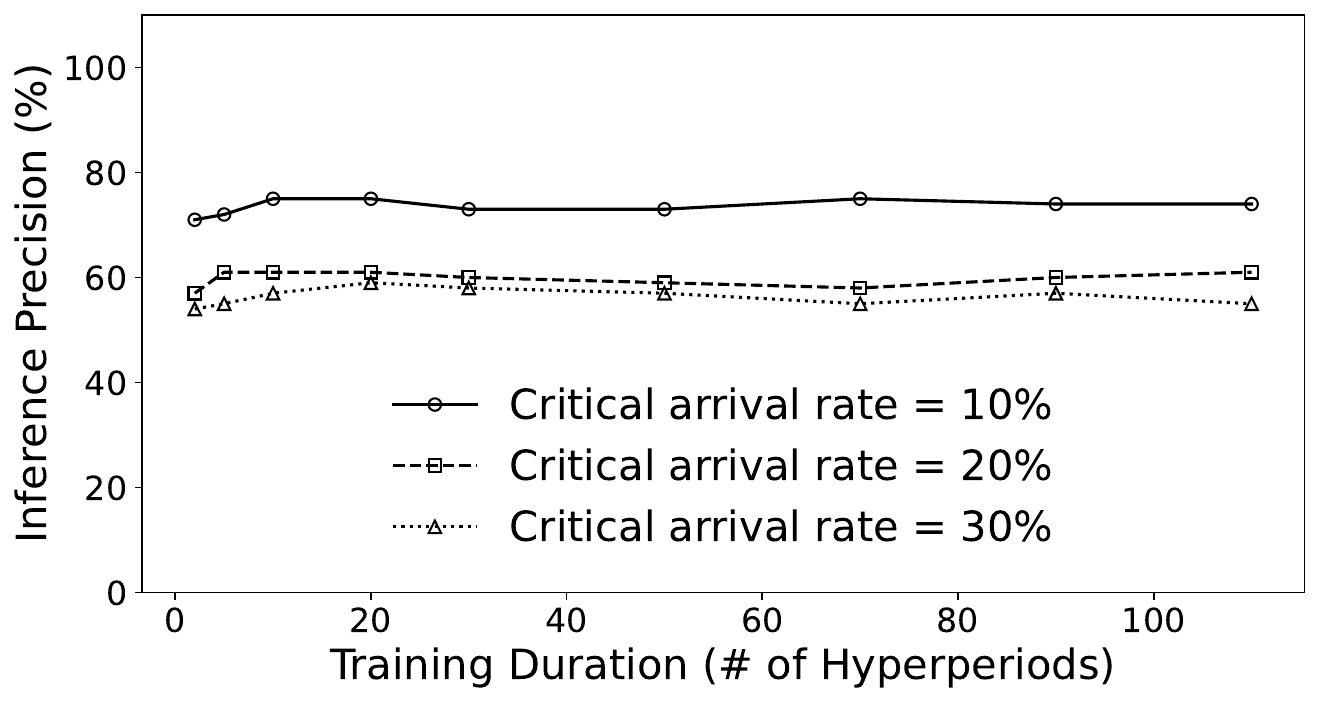}
        \caption{Training duration vs inference precision.}
        \label{fig:exp3_reg}
    \end{subfigure}
    \hfill
    \begin{subfigure}{0.49\linewidth}
        \centering
        \includegraphics[width=\linewidth]{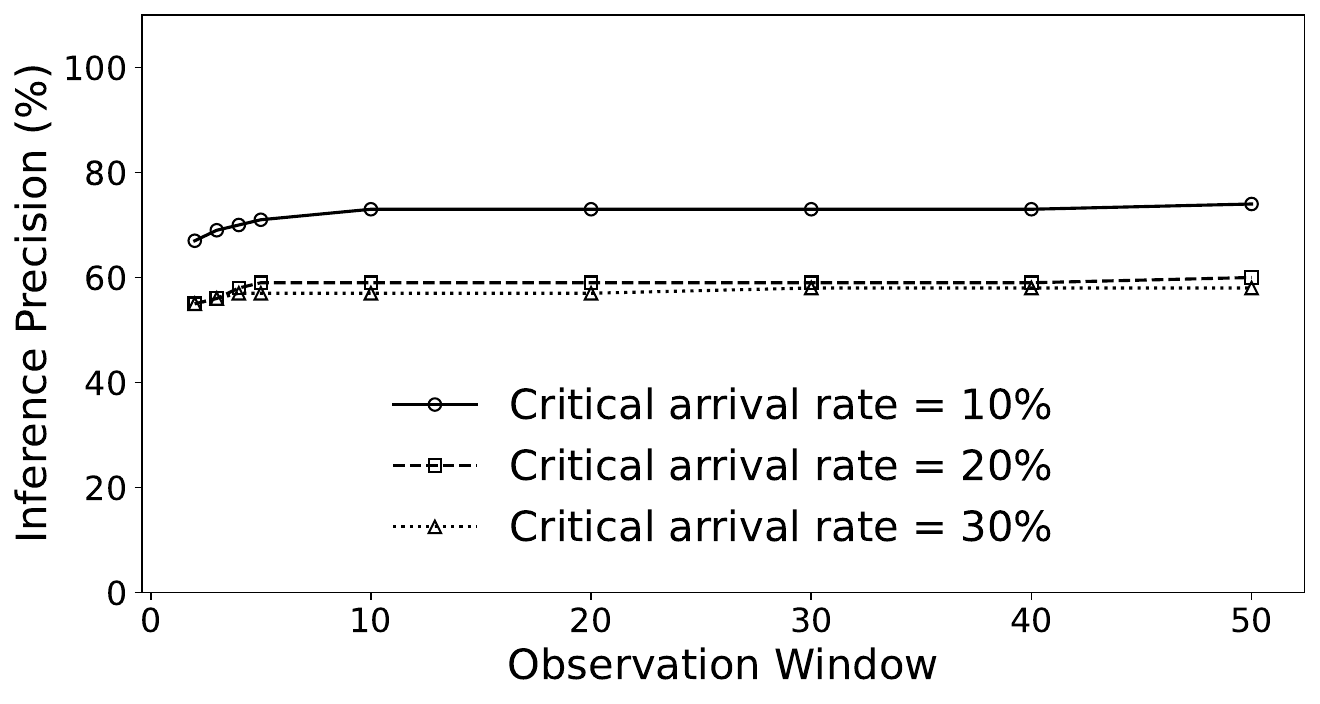}
        \caption{Observation window vs inference precision.}
        \label{fig:exp3_reg}
    \end{subfigure}
    \caption{Inference precision for ArduPilot~\cite{ardupilot} taskset. The findings are similar to those we observed using synthetic parameters (Fig.~\ref{fig:ip_combined}).}
    \label{fig:ap_combined_history}
\end{figure}

\begin{figure}[!t]
    \centering
    \begin{subfigure}{0.49\linewidth}
        \centering
        \includegraphics[width=\linewidth]{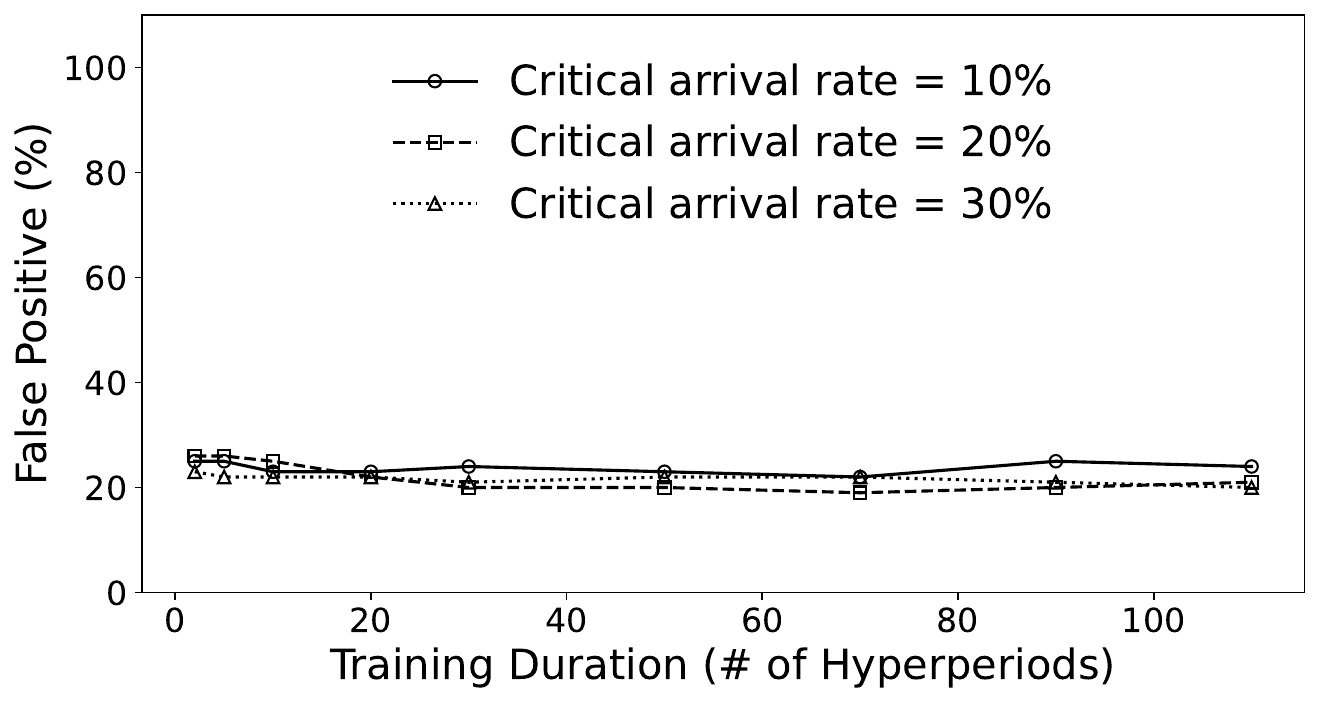}
        \caption{Training duration vs false positive.}
        \label{fig:exp3_fp}
    \end{subfigure}
    \hfill
    \begin{subfigure}{0.49\linewidth}
        \centering
        \includegraphics[width=\linewidth]{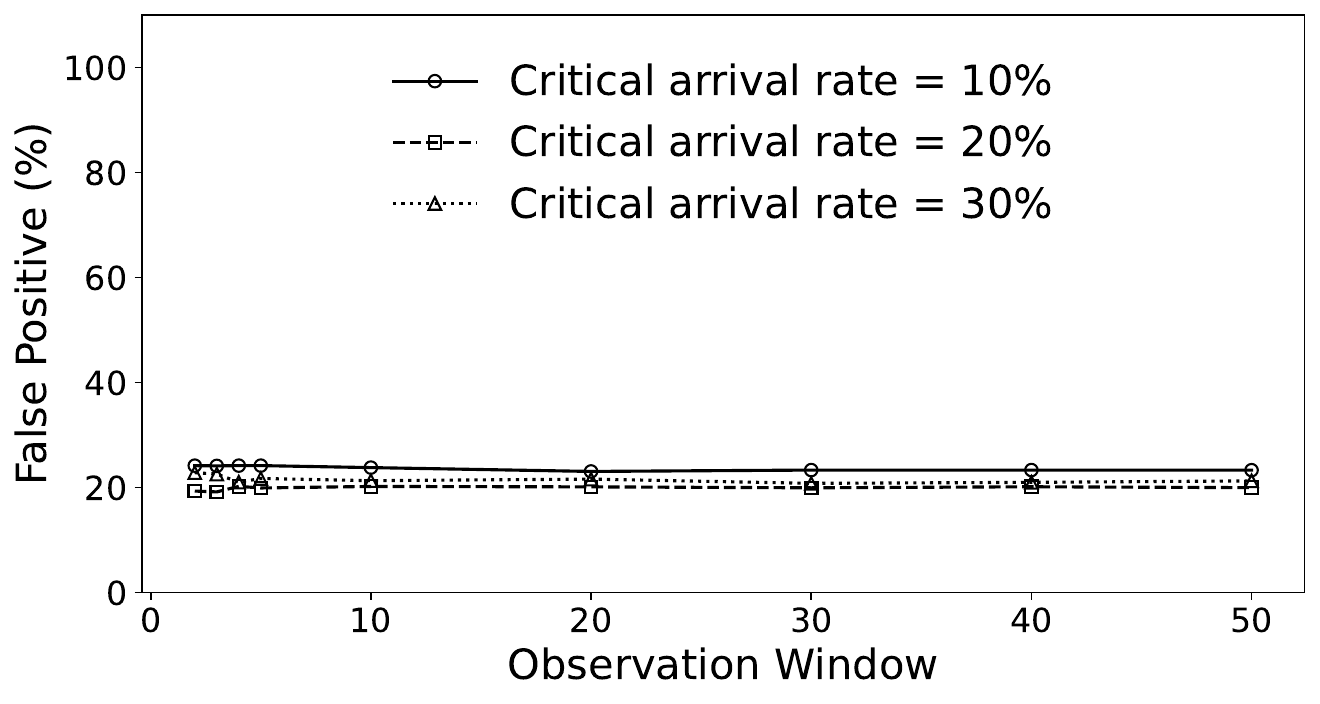}
        \caption{Observation window vs false positive.}
        \label{fig:ap_fp_history}
    \end{subfigure}
    \caption{False positive rates for ArduPilot~\cite{ardupilot} taskset. The findings are similar to those we observed using synthetic parameters (Fig.~\ref{fig:fp_combined}).}
    \label{fig:exp3_combined}
\end{figure}

The experiments conducted above use synthetically generated tasksets with commonly used parameters for broader design-space exploration. In the follow-up experiments, we want to analyze information leakage and inference performance on a realistic taskset. For this, we consider a widely-used UAV autopilot system, ArduPilot~\cite{ardupilot}. The ArduPilot UAV controller has 18 real-time tasks (defined in \texttt{/ArduCopter/Copter.cpp}). For demonstration purposes, we assume a critical task \texttt{AP\_Camera} that is used for the UAV's vision system as the \textit{victim} task and a low-priority \texttt{check\_dynamic\_flight} that updates navigation logic based on velocity as the \textit{observer} task. In Fig.~\ref{fig:ap_combined_history} and Fig.~\ref{fig:exp3_combined}, we carried out experiments similar to Fig.~\ref{fig:ip_combined} and Fig.~\ref{fig:fp_combined} but using taskset parameters that are used in ArduPilot. We find similar findings that we observed in synthetic taskset parameters. If an attacker were to perform inference on critical job arrivals for the ArduPilot autopilot system, the inference precision ranges from $50\%$ to $70\%$, and false positive rates are less than $25\%$.

\begin{center}
\fbox{\begin{minipage}{0.97\linewidth}
\itshape
We achieved less than $25\%$ false positives to correctly infer critical job arrivals for a taskset that is used in the ArduPliot system.
\end{minipage}}
\end{center}

\subsection{Overhead Analysis}\label{sec:overhead_pi}

In following set of experiments, we measure the timing and memory overheads in building the statistical model as well as the extra overheads associated with runtime inference. We conducted our experiments on a Raspberry Pi 4 Model B with 4 GB of RAM. Our implementation was done in C as this is typically used for coding real-time tasks.

\begin{figure}[!t]
    \centering
    \begin{subfigure}{0.49\linewidth}
        \centering
        \includegraphics[width=\linewidth]{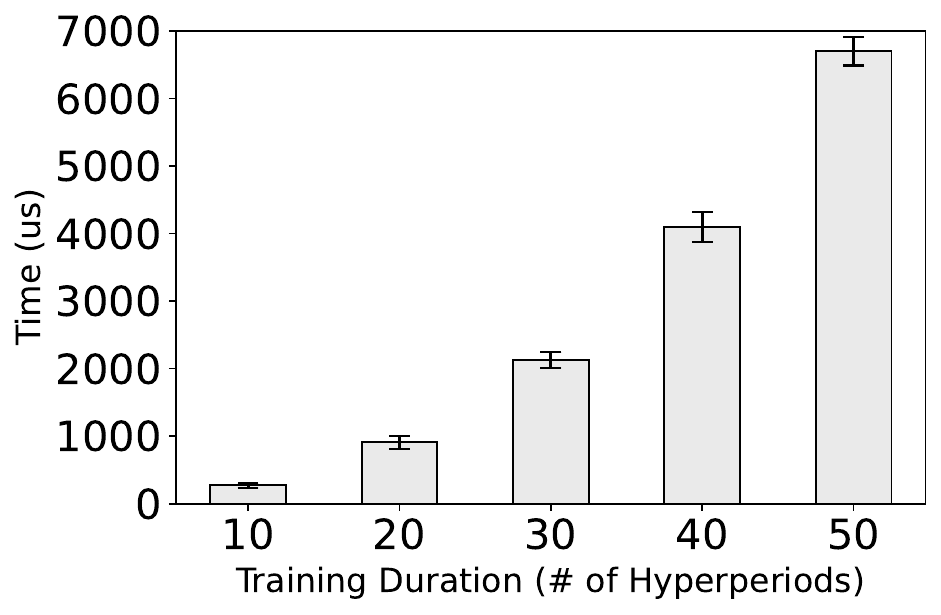}
        \caption{Model building time (done offline).}
        \label{fig:time_build}
    \end{subfigure}
    \hfill
    \begin{subfigure}{0.48\linewidth}
        \centering
        \includegraphics[width=\linewidth]{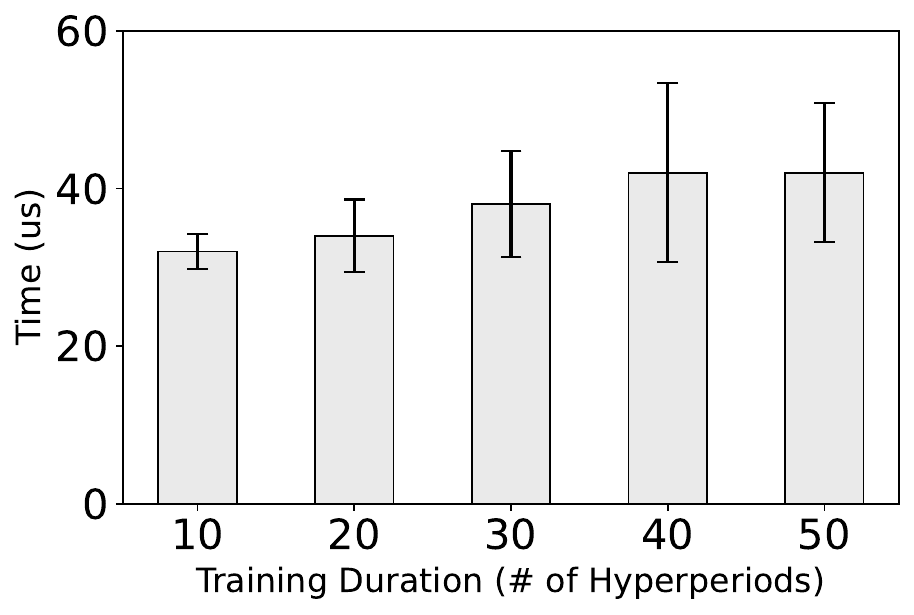}
        \caption{Inference time (runtime prediction).}
        \label{fig:time_pred}
    \end{subfigure}
    \caption{Timing overheads for building the model (left) and performing inference (right). The runtime inference overheads are relatively low (around 50 ms).}
    \label{fig:time_pi}
\end{figure}

\paragraph*{Timing Overhead}

In Fig.~\ref{fig:time_build}, we show the time it takes to build the model (PST + clustering) for different training durations. Likewise, Fig.~\ref{fig:time_pred} shows the inference overhead. We repeated the experiments 100 times and reported the $90$\textsuperscript{th} percentile time. As expected, a larger training duration increases training time as the model needs to parse more data to build the tree and form the clusters. Even though we run the experiments on an embedded platform, this cost is offline, and general-purpose computers can also be used (recall: we follow the non-interference model, and the training can be done even before system deployment by simply knowing taskset parameters). The inference timing overheads, as shown in Fig.~\ref{fig:time_pred}, are crucial as $\tau_o$ needs to make the prediction runtime from the saved model and by utilizing its past $|\mathcal{R}_o|$ response time measurements. We find that the overheads are relatively low: $\sim\!50$~us for 50 hyperperiod training duration on the Raspberry Pi board. Further recall from our earlier experiments (Fig.~\ref{fig:ip_combined} and Fig.~\ref{fig:fp_combined}) that the observer task does not need a longer training period to make an inference.

\begin{figure}[!t]
    \centering
    \begin{subfigure}{0.49\linewidth}
        \centering
        \includegraphics[width=\linewidth]{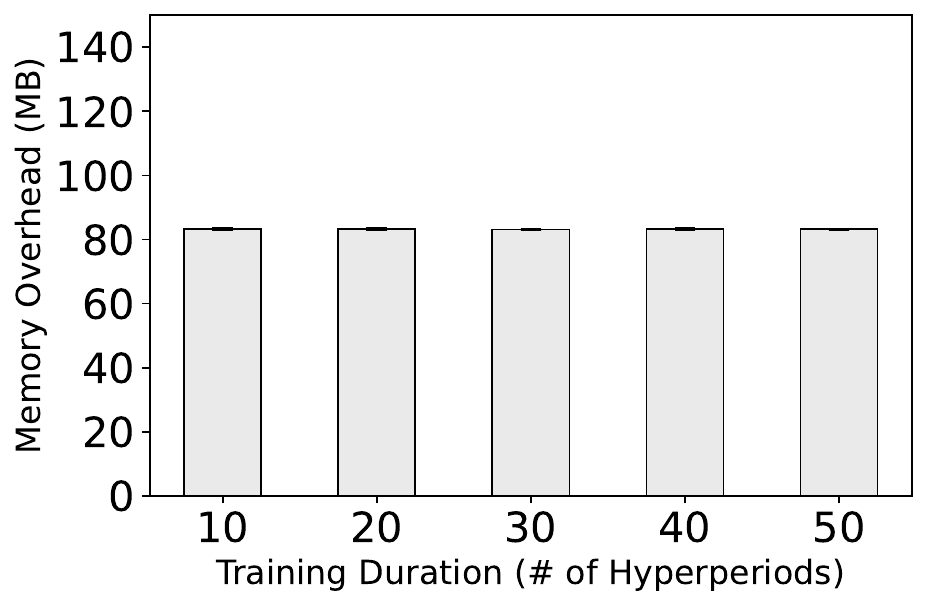}
        \caption{Memory usage during building the model.}
        \label{fig:memory_build}
    \end{subfigure}
    \hfill
    \begin{subfigure}{0.49\linewidth}
        \centering
        \includegraphics[width=\linewidth]{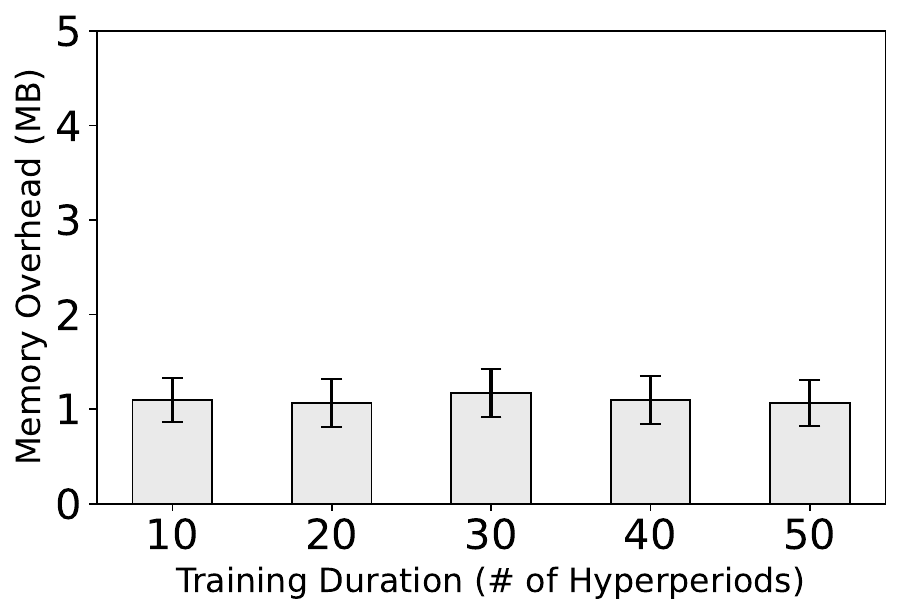}
        \caption{Memory usage during runtime prediction.}
        \label{fig:memory_build_pred}
    \end{subfigure}
    \caption{Memory overheads for building the model (left) and performing inference (right). The memory overhead for runtime inference is very low (1 MB).}
    \label{fig:memory_pi}
\end{figure}

\paragraph*{Memory Overhead}

We also carried out similar experiments to measure the memory overhead (see Fig~\ref{fig:memory_pi}). We used Linux's \verb|/usr/bin/time| utility with \verb|-f "%M"| flag to measure the peak memory usage during $\tau_o$'s code execution. As the figures show, the memory footprint remains unchanged with training duration. This is due to the fact that we only need to store a few metadata (\eg probability values) in the tree and parse it during inference, which is not a memory-hungry operation. Also, we find that runtime memory consumption is very minimal (about 1 MB). 

\begin{center}
\fbox{\begin{minipage}{0.97\linewidth}
\itshape
The runtime overheads (both in terms of timing and memory usage) for the proposed technique is relatively low, \eg around 20 ms and 1 MB for a training duration of 20 hyperperiods. 
\end{minipage}}

\end{center}

\section{Demonstration on Cyber-Physical Platforms} \label{sec:cps_demo}

Recall that the goal of our study is to assess information leakage in data-flow driven real-time systems where a task may execute in two modes depending on application requirements. How the inference information (\eg critical job of the victim) can be leveraged or exploited by the adversary is not the key focus of our work.\footnote{For instance, researchers show how target job information can be used as an attack vector~\cite{scheduleak,liu2019leaking}.} However, to further demonstrate the practicality of our analysis and how the leaked information can be used by an adversary, we carried out experiments on two cyber-physical platforms. Our first platform is a three degree-of-freedom (DoF) robot arm (Section~\ref{sec:case_study_robot}), and the second platform is a custom-built surveillance system (Section~\ref{sec:case_study_motion}).

\subsection{Demonstration 1: Manufacturing Robot}\label{sec:case_study_robot}

\begin{figure}[t]
    \centering
    \includegraphics[scale=0.25]{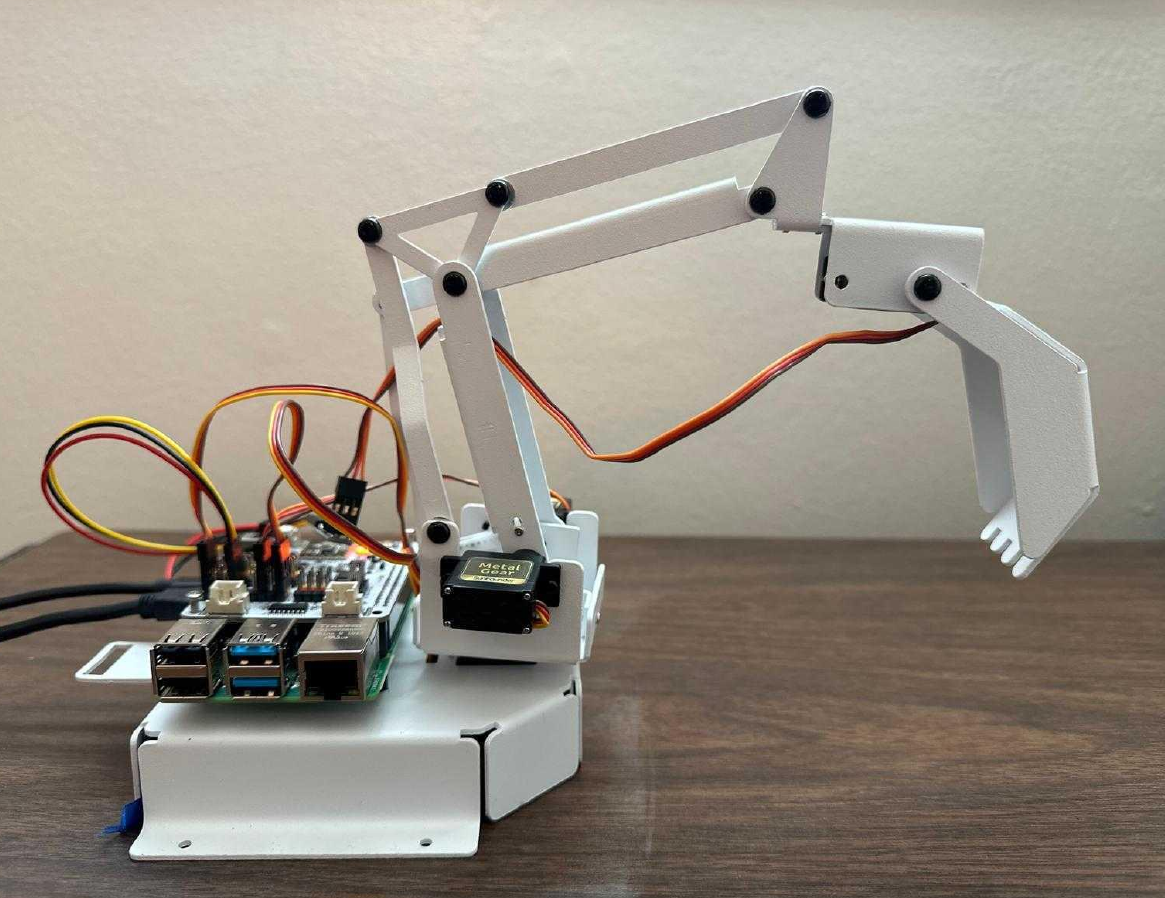}
    \caption{3-DoF robot used in our demonstration.}
    \label{fig:piarm}
\end{figure}

We use an off-the-shelf manufacturing robot (known as PiArm, manufactured by SunFounder~\cite{sunfounder_piarm}) as one of our demonstration platforms. Figure~\ref{fig:piarm} shows the robot housed in our lab. The robot is controlled by a Raspberry Pi 4 embedded board. The robot's arm movement is managed by four servos, each connected to a designated \texttt{channel} (I/O port) on a draughtboard (Adafruit motor shield~\cite{adafruitAdafruitMotor}).  We used an open-source Python-based robot controller~\cite{sunfounder_robot_hat_github} to move the robot. 

In our setup, we deploy three tasks: a high-priority victim ($\tau_v$), an intermediate task ($\tau_n$), and the observer task ($\tau_o$). The typical operation of the victim task moves its arm and shovel, \ie it performs the following actions: \texttt{move arm down}, \texttt{move shovel up}, \texttt{move arm up}, \texttt{move shovel down}. We consider a scenario where critical operation requires the robot to move its base. Hence, during the critical mode operation, the victim performs the following actions: \texttt{move base left}, \texttt{move arm down}, \texttt{move shovel up}, \texttt{move base right}, \texttt{move arm up}, \texttt{move shovel down}. Each action takes a \texttt{(channel, angle)} pair that controls the rotation of the corresponding servo to the desired angle.  We run the systems over 50 hyperperiods to take response time data for $\tau_o$ and use this data to construct our inference model. The observer task $\tau_o$ attempts to predict the arrival of the critical mode of $\tau_v$ using our learning model. When \(\tau_o\) predicts the arrival of a critical job from \(\tau_v\), it causes the corresponding servo channel to freeze by sending an arbitrary angle value. As a result, the entire robot's operation flow is broken. 

Consider this robot is part of a manufacturing assembly chain. Using the inference model introduced in this paper, an adversary can gauge---with finer precision---when a robot arm is about to pick an object from the conveyor belt. If the robot arm control task is frozen and the robot cannot pick up or move the object from the belt, the entire assembly line may collapse!

\begin{figure}[!t]
    \centering
    \begin{subfigure}{0.49\linewidth}
        \centering
        \includegraphics[width=\linewidth, height=3cm]{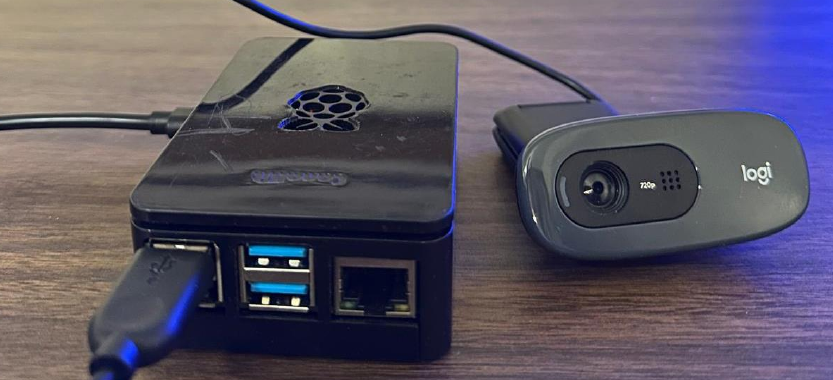}
        \caption{Raspberry Pi equipped with USB camera.}
        \label{fig:motion_pi}
    \end{subfigure}
     \hfill
    \begin{subfigure}{0.49\linewidth}
        \centering
        \includegraphics[width=\linewidth, height=3cm]{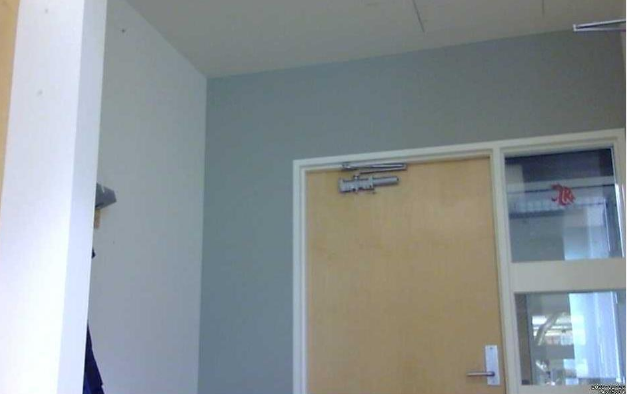}
        \caption{Images captured by the camera.}
        \label{fig:motion_image}
    \end{subfigure}
    \caption{A custom-built surveillance system used for demonstration.}
    \label{fig:motion}   
\end{figure}

\subsection{Demonstration 2: Surveillance System}\label{sec:case_study_motion}

Surveillance systems often require continuous monitoring, which leads to high energy and storage consumption. To optimize resource usage, motion-triggered recording can significantly reduce redundant data storage and energy consumption while ensuring that critical events are captured or proper actuations are performed when needed. In this case study, we consider a surveillance system that 
detects the motion in front of an entrance. We built the system using Raspberry Pi 4 Model B connected to a Logitech USB camera. Figure~\ref{fig:motion_pi} shows the system setup, and Fig.~\ref{fig:motion_image} displays a snap captured by the camera.  By default (\ie in typical mode), the system captures images in coarse granularity, and when a motion is detected (\ie critical mode), it captures fine-grained, higher-resolution images. To create this setup, we use a video analysis application, Motion~\cite{motion_project_2021} (version 4.5.1). Our system comprises five tasks, including the motion detection task. We targeted the motion detection task as our victim task ($\tau_v$). We also created three synthetic tasks ($\tau_2, \tau_3, \tau_4$) in addition to the observer task ($\tau_o$) running the inference. When no motion is detected, $\tau_v$ runs in typical mode, capturing images at 240p in each invocation. Upon detecting a motion near the entrance, $\tau_v$ immediately switches to critical mode, records a 10-second video in 720p resolution, and saves it to the filesystem. Like before, we ran the system for 50 hyperperiods to build the model for inference. In this setup, the false positives for correctly inferring critical jobs (\ie when motion was detected) was 16.45\%.

Consider this motion-triggered system is coupled with an actuation process (\eg the door is opened automatically when a motion is detected). If an adversary can infer when the critical task is about to run with finer precision (since we have a low false positive rate of 16.45\%) and prevent the actuation (say, opening the door), it may lock people in a facility and disrupt the normal operation of the automation system!

\section{Related Work}






Of late, the real-time community has started to explore cybersecurity issues~\cite{hasan2024sok}. However, there is a very limited exploration of information leakage.  Mohan \etal\cite{mohan2014real} introduce a framework that identifies and enforces security restrictions between tasks within real-time systems to prevent sensitive information from being leaked via storage channels. To minimize the risk of sensitive information exposure, the {\it flush task} method cleaned up shared resources before scheduling security-sensitive tasks. The authors then extended their ideas to a generalized model~\cite{Rodolfo2015,mohan2016integrating}. Building on this concept, Kang \etal\cite{Kang2022} develop a flush task-aware bin-packing algorithm that optimizes task allocation to multiple processors. While they are proactive defense mechanisms, \ca none formally analyze information leakage, and \cb they do not work for data-driven non-deterministic task models.

RedZone~\cite{mhhredzone} uses the typical execution behavior of real-time tasks to define temporal boundaries and use this information to detect the off-nominal temporal behavior. In contrast to this work, RedZone focuses on anomaly detection. Apart from security-centric exploration, a similar execution model (long and short executions) is used in literature to define constraints for under-specified tasks within onboard software~\cite{hammadeh_et_al:LIPIcs.ECRTS.2017.17}. 

FrameLeaker~\cite{isroc_mfb} allows a pair of tasks to establish a covert channel. As a result, a low-priority task to infer the execution patterns (frames) of a high-priority task. In a similar direction, Son~\etal~\cite{covert_channel_son_paper} explore timing covert channels for rate monotonic schedulers.  Volp~\etal~\cite{volp2008avoiding} study covert channels between different priorities of real-time tasks and proposed solutions to avoid
such covert channels. There exists other work~\cite{confidentiality_real_time,mitigating_sca,info_sca,kwak2019covert,rw_1,anis_workshop_paper} for discovering side and covert channel leakage.  Unlike ours, the above studies do not address the prediction of future task arrivals or non-deterministic dual-mode task arrivals.
ScheduLeak attack~\cite{scheduleak} discovers a new side channel to infer critical timing information. However, ScheduLeak assumes a fixed execution time for each job, which is orthogonal to the problem we investigate in this work.



The proposed research developed a systematic model using statistical tools to predict the arrival of a targeted task. As our demonstrations show (Section~\ref{sec:cps_demo}), such leaked information can be utilized to launch other successful attacks. To our knowledge, this paper is one of the first explorations to study information leakage for dual-mode real-time workloads.

\section{Conclusion}

Our research analyzes information leakage in data flow-driven real-time systems. We show that despite irregular arrival patterns, it is possible for a low-priority task to infer critical job arrivals of a targeted high-priority task. Our findings can further emphasize the need for system designers to examine stealthy leakage in real-time systems and explore new defense mechanisms to mitigate such unwanted information flows. 


\bibliographystyle{ieeetr}
\bibliography{bib}
\end{document}